\documentclass[twocolumn, amssymb, nobibnotes, aps, prd, nofootinbib, longbibliography]{revtex4-2}

\usepackage{amsmath}
\usepackage{amsfonts}
\usepackage{amssymb}
\usepackage{booktabs}
\usepackage[shortlabels]{enumitem}
\usepackage[T1]{fontenc}
\usepackage{graphicx}
\usepackage[colorlinks=true, linkcolor=blue, urlcolor=blue, citecolor=blue]{hyperref}
\usepackage{siunitx}
\usepackage{tikz}
\usepackage{xcolor}

\usetikzlibrary{shapes.geometric, arrows}
\tikzstyle{line}=[draw] 

\definecolor{theblue}{HTML}{95D0FC}
\definecolor{theorange}{HTML}{FBDD7E}
\definecolor{thepale}{HTML}{FEFFCA}

\newcommand{\updated}[1]{{#1}}

\newcommand{\A}{\mathcal{A}}
\newcommand{\Bcal}{\mathcal{B}^{*}_{\textrm{S}/\textrm{N}}}
\newcommand{\abs}[1]{\left|#1\right|}
\newcommand{\BSG}{B_{\textrm{S}/\textrm{G}}}

\newcommand{\dl}{\Delta \lambda}
\newcommand{\depth}{\mathcal{D}}
\newcommand{\deff}{\mathcal{D}_{\textrm{Eff}}}
\newcommand{\eah}{\texttt{Einstein@Home}}
\newcommand{\F}{\mathcal{F}}
\newcommand{\falcon}{\texttt{Falcon}}
\newcommand{\gauss}{\textrm{Gauss}}

\newcommand{\hyp}[1]{\mathcal{H}_{\textrm{#1}}}
\newcommand{\logB}{\ln{\Bcal}}
\newcommand{\Pro}{\mathrm{P}}
\newcommand{\ProS}{\Pro(\twiceFs|\hyp{S})}
\newcommand{\ProN}{\Pro(\twiceFs|\hyp{N})}
\newcommand{\mis}{m}
\newcommand{\mean}[1]{\mu_{\textrm{#1}}}
\newcommand{\N}{\mathcal{N}}
\newcommand{\Nseg}{N_{\textrm{seg}}}

\newcommand{\sqrtg}[1]{\sqrt{g\left(#1\right)}}
\newcommand{\std}[1]{\sigma_{\textrm{#1}}}
\newcommand{\Tcoh}{T_{\text{coh}}}
\newcommand{\Tobs}{T_{\text{obs}}}
\newcommand{\twiceF}{2\tilde{\F}}
\newcommand{\twiceFs}{2\tilde{\F}^{*}}
\newcommand{\Tsft}{T_{\text{SFT}}}
\newcommand{\viterbi}{\texttt{Viterbi}}
\newcommand{\x}[1]{\xi_{\textrm{#1}}}

\begin{document}

\title{Application of a hierarchical MCMC follow-up to Advanced LIGO\\ continuous gravitational-wave candidates}

\author{Rodrigo Tenorio} 
\email{rodrigo.tenorio@ligo.org}
\affiliation{Departament de F\'isica, Institut d'Aplicacions Computacionals i de Codi Comunitari (IAC3), Universitat de les Illes Balears, 
and Institut d'Estudis Espacials de Catalunya (IEEC), Carretera de Valldemossa km 7.5, E-07122 Palma, Spain}
\author{David Keitel}
\affiliation{Departament de F\'isica, Institut d'Aplicacions Computacionals i de Codi Comunitari (IAC3), Universitat de les Illes Balears, 
and Institut d'Estudis Espacials de Catalunya (IEEC), Carretera de Valldemossa km 7.5, E-07122 Palma, Spain}
\author{Alicia M. Sintes}
\affiliation{Departament de F\'isica, Institut d'Aplicacions Computacionals i de Codi Comunitari (IAC3), Universitat de les Illes Balears, 
and Institut d'Estudis Espacials de Catalunya (IEEC), Carretera de Valldemossa km 7.5, E-07122 Palma, Spain}

\date{\today}

\begin{abstract}
    We present the first application of a hierarchical Markov Chain Monte Carlo
    (MCMC) follow-up on continuous \mbox{gravitational-wave} candidates from
    \mbox{real-data} searches. \updated{The follow-up uses an MCMC sampler to
    draw parameter-space points from the posterior distribution, constructed
    using the matched-filter as a log-likelihood.} As outliers are narrowed
    down, coherence time increases, imposing more restrictive phase-evolution
    templates.  We introduce a novel Bayes factor to compare results from
    different stages: The signal hypothesis is derived from first principles,
    while the noise hypothesis uses extreme value theory to derive a background
    model.  The effectiveness of our proposal is evaluated on fake Gaussian
    data and applied to a set of 30 outliers produced by different continuous
    wave searches on O2  Advanced LIGO data. The results of our analysis
    suggest \updated{all but five outliers are inconsistent with an astrophysical
    origin under the standard continuous wave signal model. We successfully
    ascribe four of the surviving outliers to instrumental artifacts and a
    strong hardware injection present in the data.  The behavior of the fifth
    outlier suggests an instrumental origin as well, but we could not relate it
    to any known instrumental cause.}
\end{abstract}

\maketitle

\section{Introduction}
\label{sec:introduction}
Continuous gravitational waves (CWs) are persistent forms of gravitational radiation.
These \mbox{yet-to-be} detected signals are orders of magnitude weaker than compact 
binary coalescenses \cite{Abbott:2020niy}, requiring long integration times 
(months to years) to differentiate them from noise. 
\updated{Potentially detectable sources} using the current generation of ground-based 
interferometric detectors, Advanced LIGO \cite{AdvancedLIGO} and Advanced
Virgo \cite{AdvancedVirgo}, are neutron stars (NSs) presenting some non-axisymmetry 
such as crustal deformations, r-mode instabilities or free precession
\cite{Sieniawska_2019}, or the annihilation of ultralight boson clouds around 
spinning black holes \cite{PhysRevD.102.063020}.

Searching for a CW consists in filtering a data stream against a set of \updated{signal templates}, 
each of which is related to a certain set of parameters describing the CW model 
being searched for. The number of templates required to properly cover a certain parameter
space region, however, scales as a \updated{large} power of observing time \cite{Brady:1997ji}.
At a fixed computing cost, the optimal strategy is to split the data stream into
segments on which the filtering is performed, and then combine the resulting statistics
\cite{Brady:1998nj, Cutler:2005pn}. Since phase information is only fully preserved within each of these segments,
they are usually referred to as \emph{coherent segments} spanning a certain \emph{coherence time}.

The approach taken by current implementations of wide parameter space searches such as 
\cite{PhysRevD.70.082001, 2014PhRvD..90d2002A, Wette:2018bhc, 2018pas7.conf...37S, 
Dergachev:2019wqa, Covas:2019jqa} lies in the middle ground. \updated{Wide parameter
space regions are analyzed using a relatively low coherence time, ranging from
half an hour to a few weeks.} Surviving outliers are then sieved through a
suite of vetoes testing their (in)consistency with a CW signal; this includes
studying their persistence over the data stream, comparing their significance
in different detectors or checking whether they cross a frequency band
containing known instrumental artifacts \cite{SanchodelaJordana:2008dc,
PhysRevD.91.064007, Leaci:2015iuc, Zhu:2017ujz, 2019arXiv190706917M,
PhysRevD.100.024004, Abbott:2020mev}.  Other common strategies are coincidence
analyses between detectors or clustering neighbouring outliers in order to
relate them to a common cause \cite{Singh:2017kss, Beheshtipour:2020zhb,
Beheshtipour:2020nko, Tenorio:2020cqm}. Finally, if there are any surviving
outliers, various follow-up strategies use longer coherence times
\cite{Shaltev:2013kqa, Shaltev:2014toa, Papa:2016cwb, Ming:2019xse,
PhysRevLett.124.191102, Steltner:2020hfd,Abbott:2020mev}, either in a single
stage or in a hierarchical scheme where candidates are narrowed down over a
``ladder'' of coherence times.

Large-scale CW searches would benefit from a simple, general hierarchical setup, as it would allow for the systematic
follow-up of CW outliers using longer coherence times, imposing tighter constraints and reducing the presence
of outliers due to background noise.

Here we present the first complete framework to conduct hierarchical Markov Chain Monte Carlo (MCMC) follow-ups
and its application to a set of outliers obtained by different CW search pipelines on Advanced LIGO O2 data.
Our work builds on top of \cite{Ashton:2018ure}, which introduced the MCMC follow-up of CW outliers and
studied its performance on simulated signals in pure Gaussian noise.
We propose a new hypothesis test for the presence of a signal in the data after the full follow-up procedure.
The probability of the signal hypothesis is derived from first principles as proposed in \cite{coherentsemicoherentF};
the probability of the noise hypothesis is derived from the application of extreme value theory. We demonstrate the general
applicability of this follow-up strategy by analyzing outliers stemming from different analysis pipelines.

Although we restrict ourselves to outliers from CW searches for unknown \emph{isolated} sources, 
this framework and the corresponding software \cite{Keitel2021} can also be applied to outliers from searches 
for sources in binary systems \cite{PhysRevLett.124.191102, Abbott:2020mev}, glitching NSs \cite{Ashton:2018qth} 
and \mbox{long-duration} \mbox{gravitational-wave} transients \cite{Prix:2011qv, Keitel:2018pxz}. 

The paper is organized as follows: Section \ref{sec:mcmc_follow_up} describes the basic tools of CW data analysis
and overviews the application of MCMC samplers to the follow-up problem; Sec.~\ref{sec:b_star_sn} 
introduces a new statistic in terms of hypothesis testing; Sec.~\ref{sec:o2_outliers} introduces the O2 outliers 
to be analyzed and the follow-up setup. The results are presented in Sec.~\ref{sec:results}, concluding in 
Sec.~\ref{sec:conclusion}. We briefly comment on the statistical properties of the maximum 
$\F$-statistic over correlated templates in appendix~\ref{sec:appendix}.

\section{Continuous-wave data analysis: Search and follow up}
\label{sec:mcmc_follow_up}
A CW signal can be parametrized in terms of two families of parameters, namely the 
\emph{phase-evolution parameters} $\lambda$ and the \emph{amplitude parameters} $\mathcal{A}$.
This separation is motivated by the response of a GW detector to such signals 
\begin{equation}
    {h}(t;\lambda, \A) = \sum_{\mu=0}^{3} \updated{\A^{\mu}}\;h_{\mu}(t; \lambda)\;,
    \label{eq:hoft}
\end{equation}
where the functions $\A^{\mu}$ are independent of time \cite{JKS1998}.

The search for a CW signal can be stated in a Bayesian framework as a hypothesis
test between the noise hypothesis $\hyp{G}$, under which the
data consists of Gaussian noise ${n}(t)$, and the signal hypothesis 
$\hyp{S}(\lambda, \A)$, supporting the presence of a CW signal with a defined set of 
parameters within said noise ${n}(t) + {h}(t; \lambda, \A)$. The support of 
a stream of data ${x}$ for either of these hypotheses is quantified 
by the \emph{Bayes factor} \cite{jaynes_2003}
\begin{equation}
    \BSG({x}; \lambda, \A) = 
    \frac{ \Pro({x} | \hyp{S}(\lambda, \A)) }{\Pro({x} |  \hyp{G})}\;.
    \label{eq:BSG_A}
\end{equation}
Following \cite{Prix:2009tq, Whelan:2013xka}, and motivated by the linear dependency of 
Eq. \eqref{eq:hoft} on the amplitude functions $\mathcal{A}^{\mu}$, one can choose an 
appropiate set of priors $\Pro(\A)$ such that Eq. \eqref{eq:BSG_A} can be analytically
marginalized:
\begin{equation}
    \BSG({x}; \lambda) 
    = \int \mathrm{d}\A \; \BSG({x};\lambda, \A) \; \Pro(\A)
    \propto e^{\F({x}; \lambda)} \;.
    \label{eq:BSG}
\end{equation}
The statistic $\F$, which depends only on the data and the phase parameters, was originally derived 
as the maximum-likelihood estimator with respect to $\A$ \cite{JKS1998, PhysRevD.72.063006}.
This is a general detection statistic which only relies on the waveform decomposition presented
in Eq.~\eqref{eq:hoft} and hence can be applied also to variations of the CW signal model
such as sources in binaries~\cite{Leaci:2015bka} and transients~\cite{Prix:2011qv}.
Furthermore, the methods developed in this work can also be applied to CW outliers from
any kind of search using a different detection statistic, as long as they can be associated with a
parameter-space point with a certain uncertainty.

The role of Eq. \eqref{eq:BSG} is to update the prior probability on the phase evolution 
parameters $\Pro(\lambda)$ by means of the information conveyed by the data stream ${x}$. 
This can be stated in terms of Bayes' theorem as
\begin{equation}
    \Pro(\lambda | {x}, \hyp{S}) \propto \BSG({x}; \lambda) \;\Pro(\lambda)\;.
    \label{eq:posterior}
\end{equation}
We note that $\BSG$ and $\F$ have the same statistical power as they are related by a strictly monotonic
function. For the sake of later consistency, we will focus on $\F$ from now on. 
We refer the reader to \cite{Prix:2009tq, Prix:2011qv, Keitel:2013wga, Keitel:2015ova} for a more in-depth 
analysis of these statistics.

The detection problem is now stated in terms of a maximization: Given a stream of data ${x}$,
we are interested in finding the phase-evolution parameters $\lambda$ (also referred to as \emph{templates})
which maximize Eq.~\eqref{eq:posterior} or, equivalently, $\F({x}; \lambda)$. 

\subsection{Coherent and semicoherent searches}
\label{subsec:coherentsemi}

The fully-coherent $\F$-statistic can be expressed \updated{in terms of a linear filter} between the data stream and a 
signal template,
\begin{equation}
    \tilde{\F}(\lambda)
    \propto \abs{\langle {x}, {h}(\lambda) \rangle}^{2}\;,
    \label{eq:coherent_F}
\end{equation}
where $\langle\cdot\rangle$ represents a functional scalar product. Throughout this work, 
\updated{and following the convention of \cite{Prix:2012yu}, fully-coherent
quantities will be represented with a tilde; semicoherent quantities, introduced in 
Eq.~\eqref{eq:semicoherent_F}, will be represented with a caret.}
The response of $\tilde\F$ to an offset $\dl$ in the phase-evolution parameters $\lambda$ is
quantified using the \emph{mismatch} \cite{Prix:2006wm}, which can be defined in terms of a local quadratic 
approximation around the true signal parameters $\lambda$ where the mismatch has a minimum:
\begin{equation}
    \mis(\dl;\lambda) 
     = \frac{\tilde\F(\lambda) - \tilde\F(\lambda + \dl)}{\tilde\F(\lambda)} 
    \simeq \dl^{\textrm{T}} \cdot \bar{\bar{g}} \cdot \dl + \mathcal{O}(\dl^3)\;.
\end{equation}
The symmetric tensor $\bar{\bar{g}}$ is referred to as the parameter-space metric, and can be used to set
up parameter-space coverings, also known as template banks, at a certain mismatch level.
This quadratic approximation
is known to be valid up to $\mis \lesssim 0.3 - 0.5$, although latest developments on this subject suggest to 
further extend the approximation up to $\mis \sim 1$ \cite{Wette:2016raf, Allen:2019vcl, Allen:2021yuy}.

Maximizing Eq.~\eqref{eq:coherent_F} poses a computational challenge, as the number of templates to be 
considered in the optimization scales with a \updated{large} power of the total length of the data stream 
\cite{Wette:2014tca,Brady:1997ji}, while the  sensitivity only scales as the square root of it 
\cite{Wette:2011eu, Dreissigacker:2018afk}. 
As discussed in \cite{Dergachev:2010tm, Dergachev:2019wqa}, such a strong scaling stems from the tight 
restrictions imposed by the $\F$-statistic on the signal model, requiring phase coherence over the
whole duration of the data stream. 
A looser statistic can be constructed by imposing said coherence in a segment-wise manner. To do so, 
the data stream, spanning a time of $\Tobs$, is divided into $\Nseg$ segments, each of them with a duration of 
$\Tcoh$. The \emph{semicoherent} \mbox{$\F$-statistic} is then constructed by adding the coherent
\mbox{$\F$-statistics} computed in each segment
\begin{equation}
    \hat\F(\lambda) = \sum_{n=0}^{\Nseg-1} \tilde\F_{n}(\lambda)\;,
    \label{eq:semicoherent_F}
\end{equation}
\updated{
    where $\tilde\F_{n}$ refers to the coherent $\F$-statistic computed using only data within segment $n$.
This approach uncorrelates the template's phase-evolution between consecutive coherent segments, loosening
the constraints imposed on the data and widening $\F$-statistic peaks in the parameter 
space~\cite{Dergachev:2010tm, Ashton:2018ure}. In other words, given a parameter-space coordinate volume, 
the number of templates required to cover it at a given mismatch \emph{decreases} with lower $\Tcoh$. 
This implies a dependency of the parameter space metric $\bar{\bar{g}}$ on $\Tcoh$.}

The optimal strategy to sweep a wide parameter-space region under a controlled computational budget is then to use
a hierarchical scheme with a varying $\Tcoh$: The first stage surveys a parameter-space region with
\mbox{$\Tcoh \ll \Tobs$}, using an affordable number of templates. $\F$-statistic outliers are then analyzed 
with an increased coherence time, further narrowing down the \mbox{parameter-space} region of interest. 
This process continues either until \mbox{$\Tcoh = \Tobs$} or the candidate is vetoed by a complementary 
procedure \cite{Cutler:2005pn, Prix:2012yu, Papa:2016cwb, Ashton:2018ure}.

\subsection{MCMC-based follow-ups}

The follow-up of CW outliers requires to set up a template bank across the parameter-space region of interest.
Typical gridded approaches use a parameter-space metric to cover the parameter space at fixed maximum
mismatch \cite{Papa:2016cwb, Shaltev:2014toa, Shaltev:2015saa}. This approach usually requires an extensive 
campaign of software injections to be performed in order to calibrate the optimal set up in terms of sensitivity 
and computing cost~\cite{Papa:2016cwb}.

Alternatively, one could view the problem from the point of view of Bayesian inference. Eq.~\eqref{eq:posterior} 
relates the $\F$-statistic to a posterior probability distribution. This distribution can be sampled using a 
MCMC method, effectively constructing an adaptative random template bank in the
parameter space in which $\F$-statistic values will be more densely evaluated around high posterior probability
regions. As first discussed in \cite{Ashton:2018ure}, this approach achieves close to the theoretical optimal 
sensitivity for signals in Gaussian noise as long as the parameter space region is small enough to ensure a good
convergence of the MCMC.

\updated{For the purpose of estimating the effectiveness of an MCMC, as discussed in \cite{Ashton:2018ure}, 
the effective size of a parameter-space region can be computed in terms of the \emph{number of templates} 
$\N$ required to cover it at a mismatch of unity using a lattice with unit normalized thickness 
\cite{Prix:2007ks, Wette:2014tca}}
\begin{equation}
    \N\left(\Tcoh, \dl\right) = \int_{\dl} \mathrm{d} \lambda \; \sqrtg{\Tcoh} \;,
    \label{eq:N}
\end{equation}
where $g(\Tcoh)$ is the determinant of the parameter-space metric, which depends on $\Tcoh$ as explained
in Sec.~\ref{subsec:coherentsemi}, and $\dl$ represents the region being followed up.
The integral in Eq. \eqref{eq:N} must be computed along the resolved parameter-space dimensions only; 
i.e., one should \emph{not} include fractional templates, as doing so would underestimate the actual
number of templates \cite{Prix:2012yu, Cutler:2005pn, Leaci:2015bka}. For a follow-up search, the
parameter-space region under analysis is typically smaller than the scale of parameter-space correlations,
meaning $\sqrt{g}$ can be taken out of the integral as a constant and Eq.~\eqref{eq:N} simplifies to
\begin{equation}
    \N\left(\Tcoh, \dl\right) \simeq \sqrtg{\Tcoh} \; \textrm{Vol}\left(\dl\right)
\end{equation}
where $\textrm{Vol}\left(\dl\right)$ is the coordinate volume of the region being followed up.
Seminal analyses in \cite{Ashton:2018ure} and follow-up searches performed in
\cite{PhysRevLett.124.191102,Abbott:2020mev} suggest that values up to $\N^{*} \simeq 10^{3-4}$ are compatible with
effective MCMC runs in terms of convergence.

CW outliers are identified as a parameter-space point carrying an uncertainty which depends on the pipeline
used to conduct the search. Upon entering the follow-up pipeline, these uncertainties are converted into 
prior probability distributions to start the MCMC sampling. \updated{Ref.}~\cite{Ashton:2018ure} proposed the 
use of bounded uniform priors in order to restrict the surveyed parameter-space region; however, such hard boundaries
may prevent the successful follow-up of CW candidates whose parameters are shifted due to the presence
of parameter-space correlations. We propose the use of uncorrelated Gaussian priors, which concentrate their
probability density around a characteristic region while being unbounded. See Sec.~\ref{sec:results} for
details on the choice of Gaussian priors.

An MCMC-based follow-up is implemented in the \texttt{PyFstat} package \cite{Keitel2021} using the parallel-tempered
ensemble MCMC sampler \texttt{ptemcee} \cite{2013PASP..125..306F, 2016MNRAS.455.1919V} to sample the posterior 
distribution Eq.~\eqref{eq:posterior} using either the coherent (Eq.~\ref{eq:coherent_F}) or
semicoherent (Eq.~\ref{eq:semicoherent_F}) $\F$-statistic. We refer the reader to \cite{Ashton:2018ure} for an
extended discussion on the characteristics of this particular MCMC implementation. The analyses presented in this
work were performed using \texttt{PyFstat} version 1.11.3 \cite{ashton_gregory_2021_4542822}.

\subsection{A coherence-time ladder}
\label{subsec:ladder}

Early setups of hierarchical schemes were based on the optimization of computing resources in order
to achieve a prescribed level of sensitivity \cite{Brady:1998nj, Cutler:2005pn}. Alternatively, if no 
computational cost model was available, software injection campaigns were used to calibrate the number of
stages \cite{Papa:2016cwb}. For the case of an MCMC-based follow-up, one can use the quantity $\N(\Tcoh, \dl)$ 
to design a hierarchical scheme by imposing the proper convergence of the MCMC run at each stage 
\cite{Ashton:2018ure}.

Suppose a wide parameter-space semicoherent search produces an interesting outlier in the parameter-space
region $\dl^{(0)}$, where the exact shape is entirely dependent on the pipeline. \updated{Round-bracketed superindices
denote different stages of the follow-up.}
To set up a first follow-up stage, we choose a coherence time $\Tcoh^{(0)}$ such that 
\mbox{$\N(\Tcoh^{(0)}, \dl^{(0)}) \lesssim \N^{*}$}, ensuring the effective parameter-space resolution is
coarse enough for the MCMC algorithm to properly converge towards the region of interest. If successful, 
the resulting parameter-space region will be narrower, \mbox{$\dl^{(1)} \leq \dl^{(0)}$}, and a
second MCMC stage using a new coherence time $\Tcoh^{(1)}$ will be applied. This procedure is repeated until 
$\Tcoh = \Tobs$ and a final fully-coherent follow-up is performed.

In \cite{Ashton:2018ure}, a simple method was proposed to find the coherence time for a stage $j$ given the
previous stage's results. The idea is to increase the coherence time as much as possible such that 
the MCMC is able to converge to the target distribution. Since this convergence can be quantified in terms of
a maximum number of templates within a region $\N^{*}$, the new coherence time $\Tcoh^{(j)}$ can be obtained
by solving 
\begin{equation}
    \N\left(\Tcoh^{(j)}, \dl^{(j)}\right) = \N^{*}\;.
    \label{eq:convergence_condition}
\end{equation}
This choice minimizes the number of stages in the scheme, reducing the overall computing cost, while ensuring the
effectiveness of the MCMC approach. The explicit dependency of Eq.~\eqref{eq:convergence_condition} on the 
parameter-space region under analysis $\dl^{(j)}$, however, hinders the construction of a complete hierarchical
scheme.

This dependency can be removed by noticing the inherent self-similarity of MCMC stages: A successful MCMC 
follow-up stage ends up with a set of samples around a prominent global maximum, the fine structure of which is 
\emph{underresolved} because of the chosen coherence time. By progressing to the next stage, this fine structure
gets resolved and the MCMC \emph{zooms in} further towards the parameter-space maximum. The setup of a
\mbox{coherence-time} ladder is simply a problem of minimizing the number of stages to reduce computing
cost while maintaining sufficiently big underresolved regions for the MCMC follow-up to properly sample the region
of interest. This condition can be simply expressed as $\N(\Tcoh^{(j)}, \dl^{(j+1)}) \simeq 1$; hence, comparing 
consecutive stages factors out the problematic dependency and the hierarchical scheme can be constructed by 
solving the recurrence 
\begin{equation}
    \N^{*} 
    \simeq 
    \updated{\frac{\N\left(\Tcoh^{(j+1)}, \dl^{(j+1)}\right)}{\N\left(\Tcoh^{(j)}, \dl^{(j+1)}\right)}}
    = \frac{\sqrtg{\Tcoh^{(j+1)}}}{\sqrtg{\Tcoh^{(j)}}}
    \label{eq:ladder}
\end{equation}
given $\Tcoh^{(0)}$ and $\N^{*}$. A numerical solver for Eq.~\eqref{eq:ladder} is included in the \texttt{PyFstat}
package \cite{Keitel2021}. Constructing the coherence-time ladder as proposed by \cite{Ashton:2018ure} makes use 
of the so-called SuperSky metric \cite{Wette:2013wza, Wette:2015lfa} to compute the parameter-space volume
element. This metric is numerically well-conditioned, but requires $\Tcoh \gtrsim 1\; \textrm{day}$.

Alternatively, we derive an equivalent \mbox{coherence-time} ladder by considering the parameter-space volume 
reduction from one stage to the next. Let us define
\begin{equation}
    \updated{\gamma^{(j+1)}} = \frac{\textrm{Vol}\left(\dl^{(j)}\right)}{\textrm{Vol}\left(\dl^{(j+1)}\right)}
    \label{eq:v_jp1}
\end{equation}
as the parameter-space volume \emph{shrinkage} from stage $j$ to stage $j+1$. In a practical application, this
quantity can be computed by comparing the volume containing a certain amount of posterior probability from 
two consecutive stages.

Eq.~\eqref{eq:convergence_condition} can now be re-expressed as
\begin{equation}
    1 = \frac{ \N(\Tcoh^{(j+1)}, \dl^{(j+1)}) } { \N(\Tcoh^{(j)}, \dl^{(j)}) }\;,
\end{equation}
and Eq.~\eqref{eq:ladder} is generalized by including Eq.~\eqref{eq:v_jp1}
\begin{equation}
    \updated{\gamma^{(j+1)}} = \frac{\sqrtg{\Tcoh^{(j+1)}}}{\sqrtg{\Tcoh^{(j)}}}\;,
    \label{eq:ladder_new}
\end{equation}
where we can recognize $\updated{\gamma^{(j+1)}}$ as a generalized version of the refinement factor 
\updated{$\gamma$} introduced in Eq. (73) of \cite{Pletsch:2010xb} to account for the template bank 
refinement from a semicoherent stage to a fully-coherent one. 
To fully recover Eq. \eqref{eq:ladder}, we simply set $\updated{\gamma^{(j+1)}} = \N^{*}$ in every 
stage $j$.

According to this derivation, constructing a coherence ladder is equivalent to imposing a ratio of posterior 
volume shrinkage. For example, choosing \mbox{$\N^{*} \simeq 10^{4}$} is equivalent to imposing an overall volume 
shrinkage of \mbox{$\updated{\gamma} \simeq 10^{4}$} (i.e.~posterior volume is a ten-thousandth fraction of the prior volume)
at each step of the ladder. As a result, the behaviour of an MCMC stage is dependent upon its capability 
to fulfill the required shrinkage rate.

\section{Evaluating the hierarchical follow-up with a Bayes factor}
\label{sec:b_star_sn}
A multi-stage MCMC follow-up analyzes CW outliers by converging towards parameter-space regions
with a high posterior probability. After each stage, coherence time is increased, breaking up
underresolved regions into smaller ones and allowing the MCMC to further narrow down the parameters
associated to the loudest outlier. We are interested in evaluating the significance of the loudest 
template resulting from the multi-stage follow-up by comparing its actual fully-coherent 
$\F$-statistic to the expected value predicted by a previous stage of the ladder. 

We construct a new Bayes factor for this comparison, using the fully-coherent $\F$-statistic of the
loudest candidate
of the MCMC, $\twiceFs$, in order to quantify the support for the presence or lack of a
CW signal in the data. Following the definition in Eq.~\eqref{eq:BSG_A},
\begin{equation}
    \logB = \ln{\frac{\ProS}{\ProN}}\;,
    \label{eq:Bcal}
\end{equation}
where the hypotheses $\hyp{S}$ and $\hyp{N}$ correspond to the presence or lack of a signal, respectively.
As discussed in Sec.~\ref{subsec:Noise}, the use of extreme value theory allows us to formulate
$\hyp{N}$ such that it is not restricted to Gaussian noise, but includes any \mbox{exponentially-bounded}
distributions with unbounded domain. The following subsections are devoted to deriving the probability
distributions under each of these hypotheses.

\subsection{Noise hypothesis}
\label{subsec:Noise}

The noise hypothesis $\hyp{N}$ ascribes the obtained value of $\twiceFs$ to pure noise.
Under the presence of Gaussian noise, the coherent $\F$-statistic follows a chi-squared distribution 
with 4 degrees of freedom\footnote{We recall for the sake of consistency with the statistics 
literature that a chi-squared distribution with $\nu$ degrees of freedom corresponds to a Gamma 
distribution with shape parameter $k=\nu/2$ and scale parameter $\theta = 2$.},
$\twiceF \sim \chi^2_{4}$. If we consider the resulting MCMC samples as a template bank $\{\lambda\}$,
it is clear that \mbox{$\twiceFs = \max_{\lambda \in \{\lambda\}} \twiceF(\lambda)$} and the 
corresponding probability distribution is that of the maximum over a certain number of
templates $n$ \cite{Abadie:2010hv}:
\begin{equation}
    \Pro(\max \twiceF) 
    = n \cdot \chi^{2}_{4}(\max \twiceF) 
    \cdot \left[\int_{0}^{\max \twiceF} \!\!\!\textrm{d}\xi \;\chi^{2}_{4}(\xi)\right]^{n-1}\;,
    \label{eq:prob_max}
\end{equation}
where $\chi^2_{4}$ denotes the probability density function.
The argument equally holds for the case of the \emph{semicoherent} $\F$-statistic; 
in that case, however, the number of degrees of freedom of the chi-squared distribution would 
be $4 \Nseg$.

By construction, the \emph{effective} number of templates in a CW template
bank is different from the actual number of templates. This is because template banks are set up
such that no \mbox{parameter-space} point is further than a certain mismatch $\mis$ from a template 
in the bank, implying a certain degree of correlation among neighbouring templates \cite{Prix:2007ks}.
The problem of estimating the effective number of templates in a template bank has not found a 
definitive solution in the CW literature. 

\updated{A common approach, see e.g.~\cite{Papa:2020vfz}, is to evaluate the template bank on several realizations 
of Gaussian noise to numerically sample the probability distribution of the loudest outlier; the effective 
number of templates is then obtained by fitting $n$ from Eq.~\eqref{eq:prob_max} to the data. 
Another approach, firstly proposed in \cite{Wette:2009uea}, splits the results of a wide parameter-space search
into disjoint partitions such that they are equivalent to different realizations of a smaller search. The
fraction of effective templates can be fitted using Eq.~\eqref{eq:prob_max} to the loudest outlier per partition, 
obtaining $n$ through extrapolation. Further developments on this method proposed a non-parametric ansatz to 
directly estimate the distribution of the loudest candidate of a search \cite{Wette:2021tbv}.}

Here we will use a solution based on extreme value theory, which describes the three possible asymptotic 
distributions followed by the maximum of $n$ independent trials according to the tail of their individual 
probability distribution. Short-scale correlated variables, such as the ones arising in the search for CW signals, 
are also covered by the theory \cite{leadbetter1983extremes}. The family of three distributions, usually referred 
to as the \emph{generalized extreme value distribution}, is parametrized by a single parameter 
\mbox{$c \in \mathbb{R}$} (aside from the location and scale parameters), and encompasses every possible 
\updated{\emph{max-stable} distribution: the maximum value of a set of random variables following a 
generalized extreme value distribution follows itself a generalized extreme value distribution of the same class, 
albeit with different parameters.}
Each of the three possible distributions is related to $c$ being positive, null or negative, 
and encloses a different set of probability distributions in its domain of attraction
\cite{beirlant2004statistics, de2006extreme, embrechts2013modelling}. 

For our CW application, we focus on the case $c=0$, also known as the \emph{Gumbel distribution}
\begin{equation}
    \textrm{Gumbel}(\xi; \mu, \sigma)
    = \frac{1}{\sigma} \exp{\left[ 
    -\left(\frac{\xi-\mu}{\sigma}\right) - e^{-\left(\frac{\xi-\mu}{\sigma}\right)}
    \right]}\;,
\end{equation}
where $\mu$ and $\sigma$ are its location and scale parameters, respectively.
The domain of attraction of this distribution comprises a variety of exponentially bounded 
distributions, including the chi-squared distribution.
A similar procedure could be carried out for the other two families $c\neq0$, including power-law and 
finite tails, if the behavior of the background noise required so.
This argument is consistent with the empirical proposal of \cite{2017PhRvD..96j2006S}.

As noted in \updated{Appendix D} of \cite{Dreissigacker:2018afk}, the presence of correlated templates renders
Eq.~\eqref{eq:prob_max} unsuitable to describe the background noise distribution of CW searches. This is because 
the family of Gumbel distributions spanned by Eq.~\eqref{eq:prob_max} as \mbox{$n \to \infty$} has a
\emph{fixed} scale parameter \mbox{$\sigma = 2$}. The inclusion of correlated templates makes the
underlying distribution deviate from a chi-squared \cite{mathai1992quadratic}, but exponential tails 
still allow the distribution of the maxima to be described by a Gumbel distribution but
with \mbox{$\sigma \neq 2$}. Further discussion on this topic is presented in Appendix \ref{sec:appendix}.

As a result, we construct $\ProN$ by fitting \emph{both} the location and scale parameters 
of a Gumbel distribution to the background distribution associated to $\twiceFs$
\begin{equation}
    \ProN 
    = \frac{1}{\std{N}} \exp{\left[ 
    -\left(\frac{\twiceFs-\mean{N}}{\std{N}}\right) - e^{-\left(\frac{\twiceFs-\mean{N}}{\std{N}}\right)}
    \right]}\;.
    \label{eq:PHN}
\end{equation}
This approach has the advantage of circumventing the computation of an effective number of templates
by directly using the asymptotic distribution, the functional form of which is robust as long as the individual
distribution tails fall off exponentially. The typical number of templates evaluated in an MCMC follow-up is
consistent with a good convergence of the maximum distribution towards a Gumbel
\cite{RePEc:spr:testjl:v:24:y:2015:i:4:p:714-733}. Further discussion on the suitable application of extreme 
value theory to evaluate the loudest outlier of a gravitational-wave search will be presented 
elsewhere~\cite{distromax}.

\begin{figure}
    \includegraphics[width=\columnwidth]{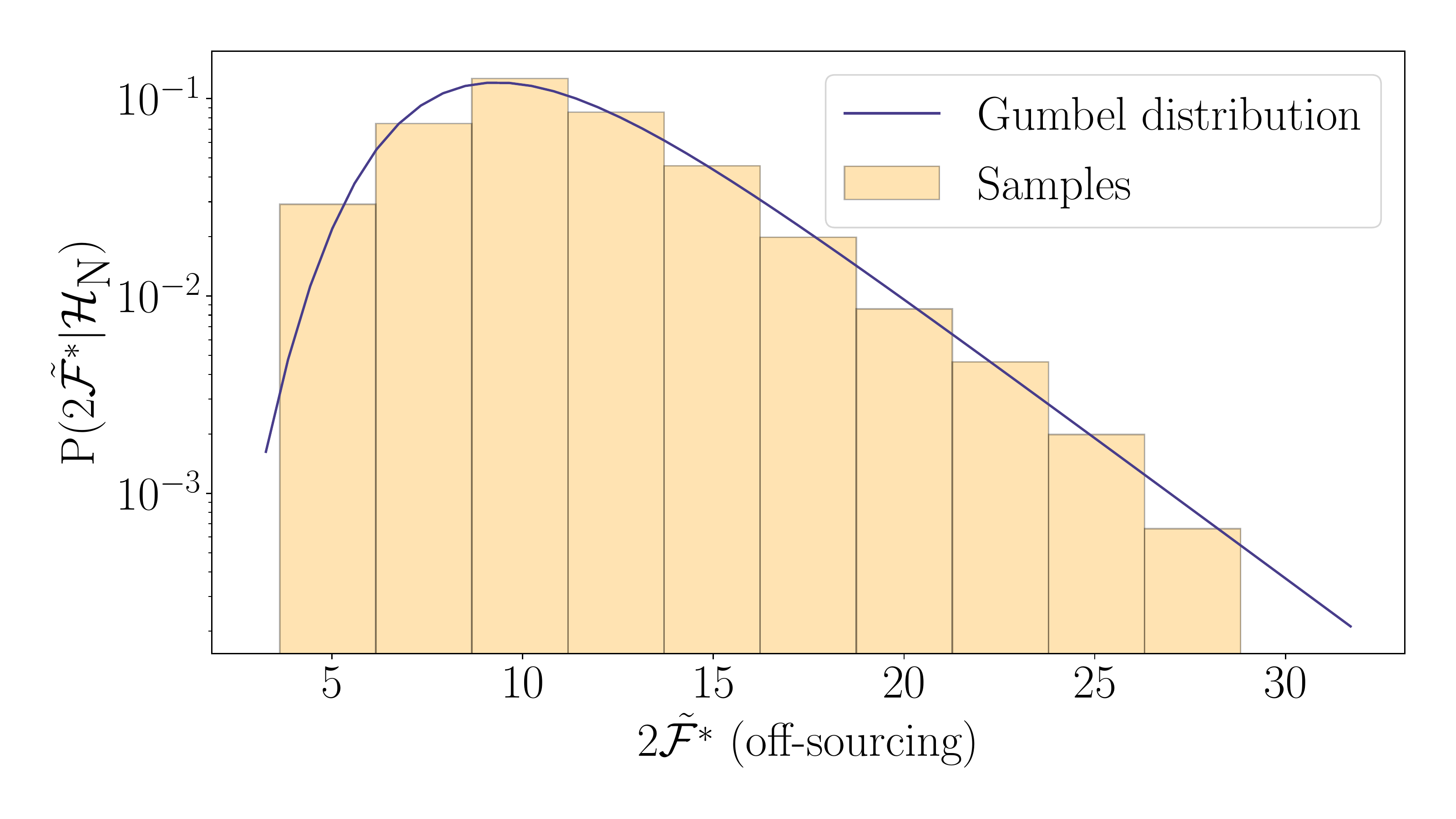}
    \caption{Distribution of the maximum $2\F$ value of a template bank obtained from its evaluation at
             \mbox{$N_{o}=600$} different off-sourced right ascensions, excluding $90^{\circ}$ around the sky
             position of the outlier of interest. The template bank corresponds to MCMC samples from the 
             fully-coherent stage follow-up of a simulated signal in Gaussian noise. The solid line represents
             the fit of a Gumbel distribution.}
    \label{fig:gumbel_offsourcing_example}
\end{figure}

To estimate the scale and location parameters of the background distribution $\mean{N}, \std{N}$, we 
apply the \emph{off-sourcing} procedure, the effectiveness of which was studied in \cite{Isi:2020uxj}. 
\updated{Off-sourcing consists in evaluating the $\F$-statistic on a template bank whose sky positions
have been purposely \emph{shifted} with respect to that of the outlier of interest. This blinds the 
detection statistic to the outlier under analysis while still sampling the same background distribution 
from the dataset. Incidentally, this takes into account template-bank correlations induced by non-Gaussian 
noise components. These correlations do not 
arise due to different templates sampling the same spectrogram data (i.~e.~overlapping \mbox{frequency-evolution}
tracks)~\cite{Wette:2009uea, Tenorio:2020cqm}, but due to the presence of correlated spectrogram data
spanning different iso-mismatch ellipsoids in the parameter space. The former kind is fundamental in the sense
that it is independent of the background; the latter is entirely dependent upon the observed data: the
wider the bandwidth of the disturbance, the lower the number of effective independent templates.}

In our concrete application, we produce \mbox{$N_{\textrm{o}}=600$} off-sourced template banks by 
randomly shifting the template's right ascension (azimuthal spherical angle), excluding a $90^{\circ}$ 
region around the sky position of interest. The declination (polar spherical angle) is unchanged in
order to maintain a constant level of sensitivity in terms of \mbox{$\F$-statistic} values. 
Fig.~\ref{fig:gumbel_offsourcing_example} shows an example of a background noise distribution obtained
through this procedure. 

The evaluation of off-sourced template banks represents the main contribution to the computing 
cost of the follow-up. The small number of outliers evaluated in this work allowed us to evaluate a set of
off-sourced samples for each of them. For the case of a large-scale follow-up, however, one could benefit from
the general properties of the Gumbel distribution to re-use a set of Gumbel parameters for different
parameter-space regions, lowering the overall computing cost.

\subsection{Signal hypothesis}
\label{subsec:signal_hypothesis}

The presence of a signal is characterized by its (squared) signal-to-noise ratio (SNR) $\rho^2$, which 
gauges the (squared) amplitude of a signal against that of the background noise 
\cite{JKS1998, PhysRevD.72.063006, PhysRevD.70.082001, Prix:2006wm}. Exact expressions for 
$\rho^2$, which include amplitude-modulation effects due to the antenna pattern of the detectors, are 
available in \cite{JKS1998, Wette:2011eu, Dreissigacker:2018afk}. 
The effect of this parameter on the probability distribution of the $\F$-statistic is to shift the 
chi-squared distribution towards a non-central chi-squared distribution,
\mbox{$2\F \sim \chi^2_{4 \Nseg}(\rho^2)$}, were the fully-coherent case corresponds to $\Nseg=1$.

As previously discussed, we are interested in comparing the consistency of $\twiceFs$ to the values
$2\hat\F^{*}$ obtained in a previous stage of the ladder. Any semicoherent stage of the ladder can be used
to construct a signal hypothesis; as discussed in more detail in Sec.~\ref{subsec:gaussian_noise}, we select
the second-to-last stage in order to benefit from the tighter constraints imposed by the signal model.
For the remainder of this section we simplify our notation by removing the asterisks, assuming every
$\F$-statistic value refers to that of the loudest candidate from the fully-coherent stage.

We construct $\Pro(\twiceF | \hyp{S})$ following the developments of 
\cite{coherentsemicoherentF}. The basic idea goes as follows: Assume a \mbox{single-template} search 
perfectly matching a signal is performed. The presence of a signal in the data, characterized 
by $\rho^2$, produces an $\F$-statistic value which depends only on $\rho^2$ and the number of coherent 
segments $\Nseg$. More specifically, obtaining a value of $2\hat{\F}$ on $\Nseg$ segments automatically 
produces an estimate on $\rho^2$, which, in turn, yields an estimation of the expected $\twiceF$ that
will be retrieved after performing a fully-coherent search.

The exact flow of information from the semicoherent to the coherent statistic can be readily expressed by
marginalizing over the unknown non-centrality parameter $\rho^2$
\begin{align}
  \Pro(\twiceF | \hyp{S})
    &= \int_{0}^{\infty} \mathrm{d} \rho^2
    \;\Pro(\twiceF | \rho^2,  2\hat{\F}, \Nseg)
    \;\Pro(\rho^2 | 2\hat{\F}, \Nseg) \nonumber\\
    &\propto \int_{0}^{\infty} \mathrm{d} \rho^2
    \;\Pro(\twiceF | \rho^2)
    \;\Pro(2\hat{\F} | \rho^2 , \Nseg)
    \;\Pro(\rho^2)\;,
    \label{eq:fc_from_semi}
\end{align}
where constant factors with respect to $2\hat{\F}$ and $\Nseg$ were omitted and the same data is being used to
compute both statistics.\footnote{\updated{This corresponds to $\kappa=1$ in the notation of \cite{coherentsemicoherentF}.}}
The choice of a prior distribution on $\rho^2$ depends on the type of search carried out; 
for a wide parameter-space search such as the ones in which we are interested it is enough to 
consider an improper uniform prior.

In going to the second line in Eq.~\eqref{eq:fc_from_semi} we have assumed no
dependency between $2\hat{\F}$ and $\twiceF$ in the sense of 
$\Pro(\twiceF | \rho^2, 2\hat{\F}, \Nseg) = \Pro(\twiceF | \rho^2)$. This relation holds exactly
if one computes each statistic on a different dataset, corresponding to the \emph{fresh data mode} in
\cite{Cutler:2005pn}.
On the other hand, if both statistics are evaluated on the same data, it represents a
\emph{conservative} choice in the sense of producing a wider distribution. This is because it neglects any
correlations between $2\hat{\F}$ and $\twiceF$. The lack of a simple way of quantifying correlations amongst 
said statistics in a general case justifies the safe approach of fresh data mode even though the same data is 
actually being used \cite{coherentsemicoherentF}.

The functional forms of the distributions in Eq.~\eqref{eq:fc_from_semi} have already been discussed in
this subsection:
\begin{equation}
    \Pro(\twiceF | \rho^2) = \chi^2_{4}(\twiceF;\rho^2)\;,
\end{equation}
\begin{equation}
    \Pro(2\hat{\F} | \rho^2 , \Nseg) = \chi^2_{4\Nseg}(2\hat{\F};\rho^2)\;.
\end{equation}
It is useful to further simplify Eq.~\eqref{eq:fc_from_semi} to a closed analytical form.
A proxy value for $\rho^2$ can be obtained by simply subtracting the expected noise-only value 
of a chi-squared distribution with $4 \Nseg$ degrees of freedom, namely 
\mbox{$\rho^2_0 = 2\hat\F - 4 \Nseg$}. Assuming $\rho^2_0 \gg 1$, chi-squared distributions can be
replaced by Gaussian distributions \cite{muirhead2005aspects, 6663749} and Eq.~\eqref{eq:fc_from_semi}
can be further replaced by a Gaussian, the peak of which corresponds to \mbox{$\mean{S}=\rho_0^{2}$}.
We refer to \cite{coherentsemicoherentF} for further details on this derivation and simply quote the
final result
\begin{equation}
    \Pro(\twiceF | 2\hat{\F}, \Nseg) = \gauss(\twiceF; \mean{S}, \std{S})\;,
    \label{eq:PHS}
\end{equation}
where
\begin{equation}
    \begin{gathered}
        \mean{S} = \rho^2_{0} \;,\\
        \std{S}^2 = 8\cdot(1 + \Nseg + \rho^2_{0})\;.
    \end{gathered}
    \label{eq:mean_std}
\end{equation}
These expressions are useful to discuss the qualitative behavior of our newly proposed Bayes factor in
different signal regimes. It will also be applicable in the analysis of software-injected signals
in Sec.~\ref{subsec:gaussian_noise}. However, due to the regime in which real-data outliers are typically found,
we do not apply this Gaussian approximation to their analysis; instead, we numerically evaluate the full version
of Eq.~\eqref{eq:fc_from_semi}.

\subsection{Bayes factor}
\label{subsec:bayes_factor}

\begin{table}
    \begin{tabular}{cc}
        \toprule
        \multicolumn{2}{c}{Parameters in Gaussian noise} \\
        \midrule
        $\mean{S}$ & $(3 - 10) \times 10^{3}$\\
        $\mean{N}$ & $10 - 20$\\
        $\std{S}$ & $30 - 80$\\
        $\std{N}$ & $2 - 3$\\
        \bottomrule
    \end{tabular}
    \caption{Typical location and scale parameters obtained from an injection campaign on
    Gaussian noise with an observing time of $\Tobs=9\;\textrm{months}$. Signal location
    and scale parameters where computed using the second-to-last stage of the coherence-time
    ladder. See Sec.~\ref{sec:results} for further details.}
    \label{table:mu_sigma}
\end{table}

We will now construct an overall Bayes factor
to compare the two hypotheses supporting the presence or lack of a signal in a given stream of data.
The distribution associated to the noise hypothesis, given in Eq.~\eqref{eq:PHN}, is constructed by
fitting the location and scale parameters of a Gumbel distribution to background data samples 
obtained through off-sourcing. The noise hypothesis can be defined in terms of said parameters,
namely \mbox{$\hyp{N} = \{\mean{N}, \std{N}\}$}, and the resulting distribution is
\begin{equation}
    \ln{\ProN} = -\left( \frac{\twiceFs-\mean{N}}{\std{N}} 
     + e^{-\left(\frac{\twiceFs-\mean{N}}{\std{N}}\right)}
     + \ln{\std{N}}
     \right)\;.
    \label{eq:logN}
\end{equation}
The signal hypothesis compares the statistical behavior of the loudest candidate across different
stages of the coherence-time ladder. We state the signal hypothesis as
\mbox{$\hyp{S} = \{\mean{S}, \std{S}\}$} and, to simplify the following discussion,
we write everything in this section using the Gaussian approximation given in Eq.~\eqref{eq:PHS}:
\begin{equation}
    \ln{\ProS} = 
    - \frac{1}{2} \left[ \left(\frac{\twiceFs-\mean{S}}{\std{S}} \right)^2 
    + \ln{2\pi\std{S}}\right]\;.
    \label{eq:logS}
\end{equation}
We note again that this approximated formula will \emph{not} be applied to real-data candidates,
as they are not located within the strong signal regime. Instead, we will then numerically evaluate
Eq.~\eqref{eq:fc_from_semi}.

\begin{figure}
    \includegraphics[width=\columnwidth]{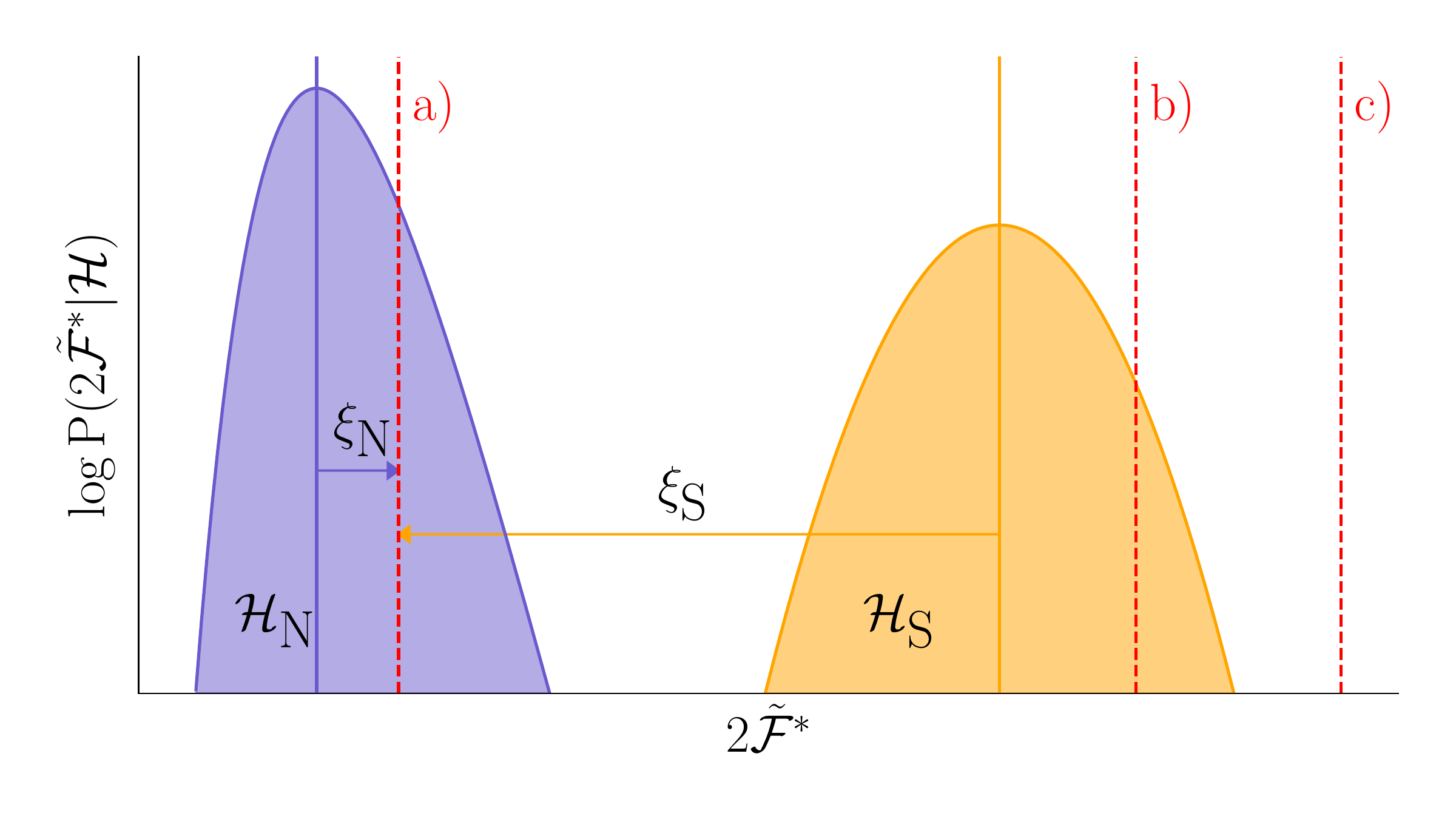}
    \caption{Illustration of different regimes in which an outlier could be located. Shaded
    regions represent probability distributions associated to the indicated hypothesis. Dashed
    vertical lines refer to the enumerated labels in the text.}
    \label{fig:xi_labeled}
\end{figure}

\begin{figure}
    \includegraphics[width=\columnwidth]{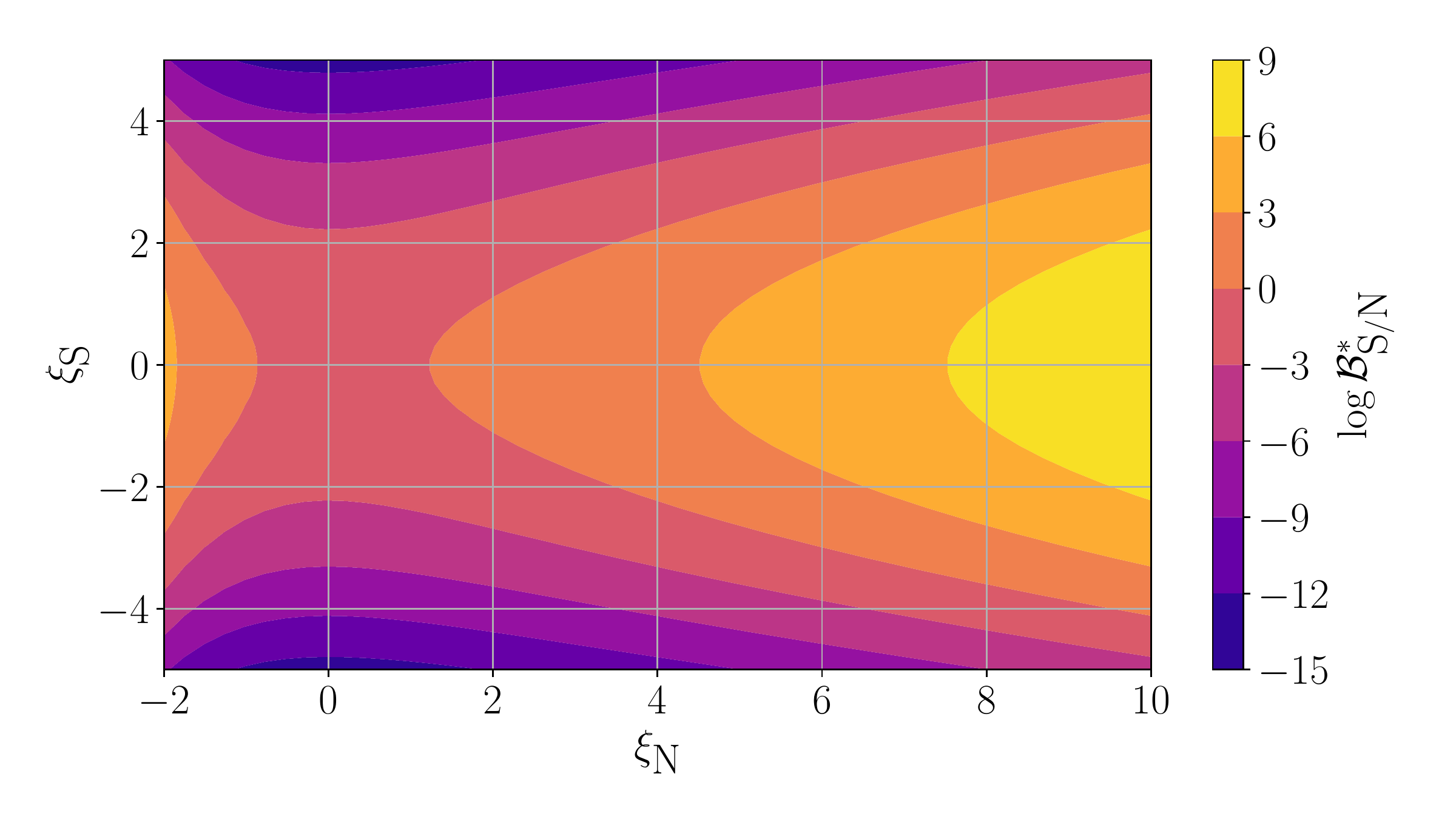}
    \caption{Bayes factor in terms of the discrepancy of an outlier with respect to the noise
    and signal hypothesis as described in Eq.~\eqref{eq:Bf}. Numerical values are computed using
    $\std{N}=3$ and $\std{S}=30$, consistent with Table~\ref{table:mu_sigma}. This representation
    will be referred to as the $(\x{N}, \x{S})$ plane.}
    \label{fig:log_Bstar_SN}
\end{figure}

It is useful to introduce the following auxiliary variables
\begin{equation}
    \x{S}= \frac{\twiceFs-\mean{S}}{\std{S}} \;, \quad \x{N} = \frac{\twiceFs-\mean{N}}{\std{N}}\;,
\end{equation}
which measure the \emph{discrepancy} of the retrieved $\twiceFs$ value with respect to the most
probable values under the signal and noise hypothesis, respectively.

Combining Eqs.~\eqref{eq:logN} and \eqref{eq:logS} we obtain an explicit expression for 
Eq.~\eqref{eq:Bcal}
\begin{equation}
    \logB = - \frac{1}{2} \xi^2_{\mathrm{S}} 
    + \xi_{\mathrm{N}} + e^{-\xi_{\mathrm{N}}} 
    + \ln{\frac{\std{N}}{\sqrt{2\pi\std{S}}}}\;.
    \label{eq:Bf}
\end{equation}
Example values of the involved quantities for the use case later in this paper are summarized 
in Table \ref{table:mu_sigma}. We proceed to analyze the general behavior of this new statistic
under different conditions.

The operating point of wide parameter-space searches is generally such that outliers being followed up
are significant enough so that $\x{N} > 0$, in the sense that a more sensitive method can be applied once
the parameter-space region has been narrowed down. It is also reasonable to expect $\mean{S} > \mean{N}$, 
although this assumption may not be valid in case of very deep searches.

We distinguish three interesting regimes of behavior of Eq.~\eqref{eq:Bf},
labeled in Fig.~\ref{fig:xi_labeled} using dashed vertical lines:
\begin{enumerate}[a)]
    \item The candidate is consistent with a noise fluctuation, returning \mbox{$\x{N} < \x{S}$},
        hence \mbox{$\logB < 0$} and the signal hypothesis is disfavored. 
    \item The candidate is consistent with the signal hypothesis \mbox{$\x{S} \sim 0$}; hence, the 
        dominant contribution to the Bayes factor is given by the discrepancy with respect 
        to the noise hypothesis \mbox{$\logB \sim \x{N}$}. This is the expected behavior of a detection
        statistic: the favoring towards the signal hypothesis is directly proportional to the discrepancy
        with respect to background noise. 
    \item The candidate is beyond the region expected by the signal hypothesis, meaning
        \mbox{$\logB \simeq -\frac{1}{2}\x{S}^2 + \x{N}$}. 
        This novel behavior is due to the chosen signal hypothesis:
        \updated{
        As opposed to the $\F$-statistic's signal hypothesis, which results in a monotonic function
        of SNR, Eq.~\eqref{eq:fc_from_semi} establishes a particular region of interest centered at 
        \mbox{$\x{S}=0$}, penalizing deviations towards \emph{both} sides of it.}
\end{enumerate}

\tikzstyle{startstop} = [rectangle, rounded corners, minimum width=3cm,
                         minimum height=1cm, text centered, draw=black, 
                         fill=red!30]
\tikzstyle{io} = [trapezium, trapezium left angle=70, trapezium right angle=110, 
                  minimum width=0.3cm, minimum height=1cm, text centered, 
                  draw=black, fill=theblue]
\tikzstyle{process} = [rectangle, minimum width=1cm, minimum height=1cm, 
                       text centered, draw=black, fill=thepale]
\tikzstyle{decision} = [diamond, minimum width=1cm, minimum height=1cm, 
                        text centered, draw=black, fill=theorange]
\tikzstyle{arrow} = [thick,->,>=stealth]

\begin{figure}

    \begin{tikzpicture}[node distance=2cm]
        \node (input) [io] {SFT Data, $\Tcoh^{(j=0)}, \dl^{(j=0)}, \N^{*}$};
        \node (mcmc) [process, below of=input] {Run MCMC};
        \node (endmcmc) [decision, below of=mcmc, yshift=-0.5cm] {$\Tcoh^{(j)}=\Tobs$?};
        \node (stagep1) [process, right of=endmcmc, xshift=2cm] {$j$ += $1$};
        \node (stagep1Tcoh) [process, right of=mcmc, align=center, xshift=2cm] 
        {Solve for $\Tcoh^{(j)}$\\$\N^{*} = \frac{\sqrtg{\Tcoh^{(j)}}}{\sqrtg{\Tcoh^{(j-1)}}}$};
        \node (retrieve) [process, below of=endmcmc, yshift=-0.5cm] 
        {Retrieve final-stage samples $\{\lambda\}$ and $\twiceFs$};
        
        \node (offsource) [process, below of=retrieve, align=center] 
        {Off-source\\sky position\\$\alpha_{i} = \alpha + \Delta \alpha_{i}$};
        \node (leftholder1) [process, left of=offsource] {\dots};
        \node (rightholder1) [process, right of=offsource] {\dots};

        \node (evaloffsource) [process, below of=offsource, align=center] 
        {Compute\\$\twiceF(\left\{\lambda_{i}\right\})$};
        \node (leftholder2) [process, left of=evaloffsource] {\dots};
        \node (rightholder2) [process, right of=evaloffsource] {\dots};
       
        \node (collect) [process, below of=evaloffsource, align=center] {Estimate $\Pro(\max \twiceF | \hyp{N})$};

        \node (PHN) [process, below of=collect, align=center] {Compute $\Pro(\twiceFs|\hyp{N})$};
        \node (PHS) [process, right of=PHN, xshift=2cm] {Compute $\Pro(\twiceFs|\hyp{S})$};
        \node (BstarSN) [io, below of=PHN, xshift=2cm] {Compute $\logB$};

        \draw [arrow] (input) -- (mcmc);
        \draw [arrow] (mcmc) -- (endmcmc);
        \draw [arrow] (endmcmc) -- node[anchor=south] {No} (stagep1);
        \draw [arrow] (endmcmc) -- node[anchor=west] {Yes} (retrieve);
        \draw [arrow] (stagep1) -- (stagep1Tcoh);
        \draw [arrow] (stagep1Tcoh) -- (mcmc);
        \draw [arrow] (retrieve) -| (PHS);
        \draw [arrow] (retrieve) -- node[anchor=west]{$i=1, \dots, N_{\textrm{o}}$} (rightholder1);
        \draw [arrow] (retrieve) -- (leftholder1);
        \draw [arrow] (retrieve) -- (offsource);
        \draw [arrow] (offsource) -- (evaloffsource);
        \draw [arrow] (leftholder1) -- (leftholder2);
        \draw [arrow] (rightholder1) -- (rightholder2);
        \draw[arrow] (evaloffsource) -- (collect);
        \draw[arrow] (leftholder2) -- (collect);
        \draw[arrow] (rightholder2) -- (collect);
        \draw[arrow] (collect) -- (PHN);
        \draw [arrow] (PHN) -- (BstarSN);
        \draw [arrow] (PHS) -- (BstarSN);
    \end{tikzpicture}
    \caption{Flowchart ilustrating the computation of $\logB$ for a CW outlier.}
    \label{fig:flowchart}
\end{figure}
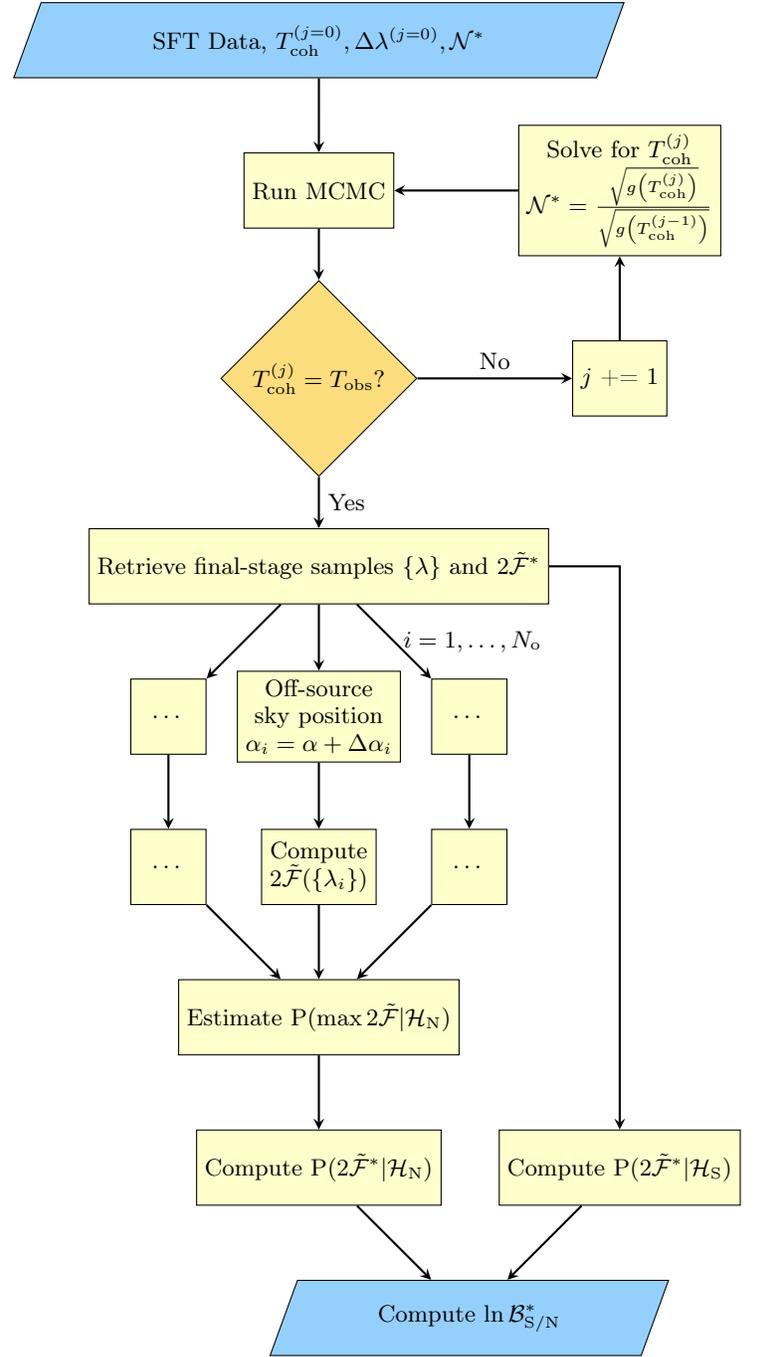
 
A complementary description of Eq.~\eqref{eq:Bf} is shown in Fig.~\ref{fig:log_Bstar_SN}, 
where $\logB$ is shown on the $(\x{N}, \x{S})$ plane. These two variables, which represent the discrepancy
of $\twiceFs$ with respect to the noise and signal hypothesis, are related by 
\begin{equation}
    \x{N} = \frac{\std{S}}{\std{N}}\x{S} + \frac{\mean{S}-\mean{N}}{\std{N}}\;,
    \label{eq:linear}
\end{equation}
meaning that once $\{\mean{S}, \std{S}\}$ and $\{\mean{N}, \std{N}\}$ are determined, the detection statistic is
restricted to a straight line in $(\x{N}, \x{S})$.
This description also clarifies the behavior of $\logB$ in case b) of Fig.~\ref{fig:xi_labeled}.
In said case, $\x{S} \sim 0$ and then \mbox{$\logB \sim (\mean{S} - \mean{N})/\std{N}$}, so that it is the
\emph{combined} action of a high SNR (\mbox{$\mean{S} > \mean{N}$}) and a low discrepancy 
with respect to the expected value according to previous stages ($\x{S} \sim 0$) what decides on the consistency 
of a CW candidate with respect to the signal or noise hypothesis.

A summary of the construction and practical computation of $\logB$ is shown as a flowchart in
Fig.~\ref{fig:flowchart}.

\section{Follow-up of outliers in LIGO O2 data}
\label{sec:o2_outliers}
We now present the first application of a multi-stage MCMC-based hierarchical follow-up
on real data by studying a set of 30 outliers obtained by different CW searches on 
Advanced LIGO O2 data. These are final-stage outliers resulting from the application of
a complete search pipeline, including a set of vetoes depending upon the particularities
of each search.

Section \ref{subsec:o2_outliers} briefly describes the main 
traits of the searches from which outliers are collected. The complete set of 
outliers to be followed up is reported in Table \ref{table:outliers}. The follow-up 
setup is described in Sec.~\ref{subsec:followup_setup}.

\subsection{Continuous-wave search outliers from O2 data}
\label{subsec:o2_outliers}

\subsubsection{All-sky \falcon{} search}

The \falcon{} pipeline \cite{Dergachev:2019wqa} is designed to survey wide parameter-space regions  
using a so-called \emph{loosely coherent} approach \cite{Dergachev:2010tm, Dergachev:2011pd, Dergachev:2018ftg},
increasing its robustness against small deviations from the standard CW signal model \cite{Dergachev:2010tm}.

We are interested in 18 outliers reported in two all-sky searches targeting two different frequency bands of the
Advanced LIGO O2 dataset: mid frequency (\mbox{$500 - 1700\;\textrm{Hz}$}) \cite{Dergachev:2020fli} and 
high frequency (\mbox{$1700 - 2000\;\textrm{Hz}$}) \cite{Dergachev:2020upb}. These searches intended to unveil 
unknown low-ellipticity sources by analyzing a restricted set of spindown rates 
(\mbox{$\abs{f_{1}} \lesssim 3 \cdot 10^{-12}\;\textrm{Hz}/\textrm{s}$}). 
These outliers are the result of a four-stage search using four different coherent times, namely 12, 24, 48 and
144 hours. After each stage, only those templates over a specified threshold were further followed up.

An additional low-frequency Falcon search was recently reported in
\cite{Dergachev:2021ujz}. As will be shown in Sec.~\ref{subsec:fu} regarding
low-frequency outliers from the other searches discussed below, the greater
number and variety of instrumental artifacts in the low-frequency data somewhat
hinder the effectiveness of this first incarnation of our follow-up method, as
they are not directly addressed by the noise hypothesis.  Therefore, we leave a
re-analysis of the new low-frequency Falcon outliers for future work.

\subsubsection{Directed \eah{} search}

\eah{} is a large-scale computing framework based on the volunteer-computing platform \texttt{BOINC} 
\cite{2019arXiv190301699A} on which the Global Correlations Transform pipeline
\cite{Pletsch:2008gc, Pletsch:2009uu, 2019PhRvD..99h2004W}, intended to perform deep (very sensitive) 
searches across wide parameter-space regions, is deployed. This pipeline is flexible enough so as to be 
reconfigured into a directed pipeline, using astrophysical information obtained by electromagnetic means 
to restrict the sky positions to search on.

We are interested in the surviving outlier from a directed search for CWs from central compact objects in three
supernova remnants \cite{Papa:2020vfz}. Said outlier is associated to the central compact object
known as \mbox{1 WGA J1713.4-949} \cite{pfeffermann1996rosat} and located in \mbox{SNR G347.3-05};
for consistency with \cite{Papa:2020vfz}, we will simply refer to it as J1713.

\updated{This outlier is a sub-threshold candidate from an earlier \eah{} search on O1 data directed towards the same 
supernova remnants \cite{Ming:2019xse}, which was then re-analyzed using O2 data. The statistical basis of 
the re-analysis was similar to the techniques explained in Sec.~\ref{subsec:signal_hypothesis}, comparing the 
significance of a candidate on different data streams with respect to the expected significance deduced from 
the initial analysis. As reported in \cite{Papa:2020vfz}, the outlier under analysis is inconsistent
with Gaussian noise, but cannot be associated to the signal hypothesis either.}

\subsubsection{Fomalhaut b \viterbi{} search}

The \viterbi{} method spans a family of search pipelines which use a Hidden Markov model (HHM) to 
describe the  frequency evolution of a CW signal
\cite{Suvorova:2016rdc, 2017PhRvD..96j2006S, 2018PhRvD..97d3013S, 2019PhRvD.100b3006B}. 
Such a signal model is able to incorporate stochastic contributions into the analysis (e.g. timing noise or
spin-wandering due to the presence of an accreting companion \cite{Mukherjee:2017qme}).

Ref.~\cite{Jones:2020htx} reports on a \viterbi{} search for CWs directed at Fomalhaut b, an astrophysical 
object whose exact nature is still surrounded by debate 
\cite{Kalas:2008ff, 2012ApJ...760L..32C, 2015MNRAS.448..376N, 2020PNAS..117.9712G}.
This search complements a previous one performed on Advanced LIGO O1 data using an $\F$-statistic search
assuming the standard deterministic evolution of a CW \cite{Abbott:2018qee}. 

The search setup assumes spindown to be the main contribution to the frequency evolution, considering timing
noise as a sub-dominant component. This is done by imposing a \emph{biased} random walk as a HMM, in the sense
that evolution towards higher frequencies is forbidden\footnote{This condition drastically reduces the space of
possible frequency evolutions contemplated by the HMM model, easing the application of a model-based pipeline
to follow up or estimate the exact parameters of any resulting candidates.}. 
The search was performed using \mbox{$\Tcoh=5\;\textrm{days}$} and surviving candidates were sieved though a set of 
consistency vetoes. In the end, a single outlier was reported for further exploration.

\subsubsection{H.E.S.S. \viterbi{} search}

Another implementation of the \viterbi{} pipeline, similar in scope and assumptions to that mentioned above,
was used to perform a search on a set of ten pulsars observed by very high-energy $\gamma$-ray surveys in
\cite{Beniwal:2021hvc}.

The search looks for CW emission at once, twice and $4/3$ of the rotational frequency of the targeted pulsars
in order to address several emission mechanisms \cite{Sieniawska_2019}.
After assessing the sub-dominant role of spin-wandering on frequency evolution, a biased random walk is 
implemented in a similar manner to \cite{Jones:2020htx}, selecting the maximum $\Tcoh$ allowed by the spindown
rate of each pulsar so that the frequency evolution is within the range of the HMM.

After applying a set of consistency vetoes, twelve outliers are reported for further exploration; we only 
considered ten of them as independent follow-up targets
since for two pairs of outliers, the corresponding prior parameter-space regions significantly overlap.

\subsection{Follow-up setup}
\label{subsec:followup_setup}

\begin{table*}
    \begin{tabular}{llccccccl}
\toprule
    Outlier ID &        Search &  $f_0$ [Hz] &  $f_1$ [Hz/s] &  $f_2$ [Hz/$\textrm{s}^2$] & $\alpha$ [rad] &  $\delta$ [rad] &  $t_{\textrm{ref}}$ [GPS] &                     Ref. \\
\midrule
    Falcon 4 &     High-frequency \falcon{} & 1891.756740 & $-8.22 \cdot 10^{-12}$ &   --- &    2.986956 &        1.005798 &                                1183375935 & \cite{Dergachev:2020upb} \\
    Falcon 5 &     High-frequency \falcon{} & 1892.991060 & $-1.08 \cdot 10^{-12}$ &   --- &   3.779161 &       -0.816273 &                                1183375935 & \cite{Dergachev:2020upb} \\
    Falcon 15 &    Mid-frequency \falcon{} &  900.218805 & $-2.20 \cdot 10^{-12}$ &  --- &   2.084418 &       -0.102264 &                                1183375935 & \cite{Dergachev:2020fli} \\
    Falcon 19 &    Mid-frequency \falcon{} &  514.148927 & $ 1.60 \cdot 10^{-12}$ &  --- & 2.170421 &        0.092501 &                                1183375935 & \cite{Dergachev:2020fli} \\
  Falcon 23 &      Mid-frequency \falcon{} & 1001.366228 & $ 4.30 \cdot 10^{-12}$ &    --- &   1.355837 &       -0.770266 &                                1183375935 & \cite{Dergachev:2020fli} \\
  Falcon 24 &      Mid-frequency \falcon{} &  676.195421 & $ 2.80 \cdot 10^{-12}$ &    --- &   3.847021 &       -0.101619 &                                1183375935 & \cite{Dergachev:2020fli} \\
  Falcon 25 &      Mid-frequency \falcon{} &  744.219166 & $ 2.40 \cdot 10^{-12}$ &    --- &   3.344985 &        0.612566 &                                1183375935 & \cite{Dergachev:2020fli} \\
  Falcon 29 &      Mid-frequency \falcon{} &  512.490814 & $ 1.20 \cdot 10^{-12}$ &    --- &   2.468975 &       -0.043050 &                                1183375935 & \cite{Dergachev:2020fli} \\
  Falcon 31 &      Mid-frequency \falcon{} &  983.151889 & $ 2.20 \cdot 10^{-12}$ &    --- &   3.561119 &        0.017979 &                                1183375935 & \cite{Dergachev:2020fli} \\
  Falcon 34 &      Mid-frequency \falcon{} &  886.880087 & $-1.60 \cdot 10^{-12}$ &    --- &   4.912788 &       -0.703498 &                                1183375935 & \cite{Dergachev:2020fli} \\
  Falcon 35 &      Mid-frequency \falcon{} &  988.373199 & $ 1.20 \cdot 10^{-12}$ &    --- &   0.981835 &        0.778338 &                                1183375935 & \cite{Dergachev:2020fli} \\
  Falcon 39 &      Mid-frequency \falcon{} &  514.291681 & $ 3.20 \cdot 10^{-12}$ &    --- &   0.569033 &       -0.128357 &                                1183375935 & \cite{Dergachev:2020fli} \\
  Falcon 40 &      Mid-frequency \falcon{} &  831.988473 & $ 4.00 \cdot 10^{-13}$ &    --- &   4.917347 &        1.160537 &                                1183375935 & \cite{Dergachev:2020fli} \\
  Falcon 41 &      Mid-frequency \falcon{} &  873.524608 & $ 4.00 \cdot 10^{-13}$ &    --- &   0.618991 &       -0.189450 &                                1183375935 & \cite{Dergachev:2020fli} \\
  Falcon 42 &      Mid-frequency \falcon{} &  895.421949 & $ 3.60 \cdot 10^{-12}$ &    --- &   5.105590 &        0.249163 &                                1183375935 & \cite{Dergachev:2020fli} \\
  Falcon 43 &      Mid-frequency \falcon{} & 1224.745666 & $-2.16 \cdot 10^{-12}$ &    --- &   1.715268 &        0.196184 &                                1183375935 & \cite{Dergachev:2020fli} \\
  Falcon 45 &      Mid-frequency  \falcon{} &  698.728032 & $-2.00 \cdot 10^{-13}$ &    --- &   4.557347 &       -0.724141 &                                1183375935 & \cite{Dergachev:2020fli} \\
  Falcon 46 &      Mid-frequency  \falcon{} & 1095.557400 & $-1.08 \cdot 10^{-12}$ &    --- &   4.354664 &       -0.260254 &                                1183375935 & \cite{Dergachev:2020fli} \\
       J1713 & \eah{} &  368.801379 & $-4.37 \cdot 10^{-9}$  &    $5.9 \cdot 10^{-19}$ &   4.509371 &       -0.695189 &              1131943508 &      \cite{Papa:2020vfz} \\
 Fomalhaut b &       Fomalhaut b \viterbi{} &  876.503400 & $-1.00 \cdot 10^{-12}$ &    --- &   6.011130 &        0.517000 &                                1167545066 &     \cite{Jones:2020htx} \\
 J0534+2200  &       H.E.S.S. \viterbi{} &   29.813738 & $-3.77 \cdot 10^{-10}$ &    --- &   1.459675 &        0.384225 &                                1164556817 &   \cite{Beniwal:2021hvc} \\
J1420--6048  &       H.E.S.S. \viterbi{} &   14.511294 & $-1.70 \cdot 10^{-11}$ &    --- &   3.753057 &       -1.061240 &                                1164556817 &   \cite{Beniwal:2021hvc} \\
J1420--6048  &       H.E.S.S. \viterbi{} &   19.515033 & $-2.30 \cdot 10^{-11}$ &    --- &   3.753057 &       -1.061240 &                                1164556817 &   \cite{Beniwal:2021hvc} \\
J1420--6048  &       H.E.S.S. \viterbi{} &   29.522611 & $-3.50 \cdot 10^{-11}$ &    --- &   3.753057 &       -1.061240 &                                1164556817 &   \cite{Beniwal:2021hvc} \\
J1718--3825  &       H.E.S.S. \viterbi{} &   17.503470 & $-3.00 \cdot 10^{-12}$ &    --- &   4.530116 &       -0.670585 &                                1164556817 &   \cite{Beniwal:2021hvc} \\
J1831--0952  &       H.E.S.S. \viterbi{} &   14.501823 & $-1.00 \cdot 10^{-12}$ &    --- &   4.850147 &       -0.172213 &                                1164556817 &   \cite{Beniwal:2021hvc} \\
J1831--0952  &       H.E.S.S. \viterbi{} &   15.401223 & $-1.00 \cdot 10^{-12}$ &    --- &   4.850147 &       -0.172213 &                                1164556817 &   \cite{Beniwal:2021hvc} \\
J1831--0952  &       H.E.S.S. \viterbi{} &   19.999146 & $-2.00 \cdot 10^{-12}$ &    --- &  4.850147 &       -0.172213 &                                1164556817 &   \cite{Beniwal:2021hvc} \\
J1849--0001  &       H.E.S.S. \viterbi{} &   26.308209 & $-9.00 \cdot 10^{-12}$ &    --- &  4.850147 &       -0.000375 &                                1164556817 &   \cite{Beniwal:2021hvc} \\
J1849--0001  &       H.E.S.S. \viterbi{} &   26.341209 & $-9.00 \cdot 10^{-12}$ &    --- &   4.850147 &       -0.000375 &                                1164556817 &   \cite{Beniwal:2021hvc} \\
\bottomrule
\end{tabular}
     \caption{CW search outliers of interest as reported by their original searches.
    H.E.S.S. \viterbi{} outliers will be further referred to by including their 
    corresponding frequency.}
    \label{table:outliers}
\end{table*}

\begin{table}
    \begin{tabular}{lc}
        \toprule
        Search & Estimated $\depth^{95\%}\; \left[\textrm{Hz}^{-1/2}\right]$ \\
        \midrule
        High-frequency \falcon{} & 55 --- 65 \\
        Mid-frequency \falcon{} & 45 --- 55 \\
        Directed \eah$^{*}$ & 80 --- 90 (75 --- 85)\\
        Fomalhaut b \viterbi{} & 45 --- 55 \\
        H.E.S.S. \viterbi{} & 45 --- 55 \\
        \bottomrule
    \end{tabular}
    \caption{Estimated ranges of 95\% efficiency sensitivity depths achieved by each 
    of the searches according to their reported results. \updated{The depth marked with an asterisk
    corresponds to a 90\% efficiency instead. Values in parentheses refer to the sensitivity
    depth achieved by the original search producing the outlier \cite{Ming:2019xse}.}}
    \label{table:estimated_depth}
\end{table}

We demonstrate the general application of an \mbox{MCMC-based} \mbox{multi-stage} follow-up to a set of real-data
outliers regardless of the pipeline producing them. To do so, outliers will be analyzed ignoring any information 
gathered  from any of the vetoes or follow-up stages reported in their respective searches.

The second Advanced LIGO observing run  \cite{O2Data, O2Datadoi} comprises nine months of data taken by 
the two Advanced LIGO detectors H1 (Hanford) and L1 (Louisiana) \cite{AdvancedLIGO}. 
The employed time segments are those with the ``all'' tag in \cite{O2DataStamps}. The dataset was divided 
into segments with a duration of $\Tsft=1800\;\textrm{s}$ in which Fourier transforms were computed as
explained in \cite{PhysRevD.100.024004}. We take the observing time to be $\Tobs=270\;\textrm{days}$ in
order to convert the number of segments of a stage $\Nseg$ to a coherence time as $\Tcoh = \Tobs / \Nseg$.

Our follow-ups are conducted assuming a CW signal model with two spindown components.
Since the second spindown component is only reported by the \eah{} search, we assume it to be compatible
with a null value for the other outliers and apply a canonical uncertainty of
$\delta f_{2} = 2 \cdot \left(\Tcoh \cdot \Tobs^2\right)^{-1}$ \cite{PhysRevD.70.082001}.
As discussed in \cite{Ashton:2017wui}, this increases the robustness of a search method against unmodeled 
physics, such as neutron star glitches, due to an increase of the available parameter-space correlations.

Table \ref{table:estimated_depth} collects the approximated sensitivity depth achieved by each search according
to their reported results. A comparison to the results in Figs. 8 and 9 of \cite{Ashton:2018ure}, which compute 
the detection efficiency of a four-stage MCMC follow-up starting at $\Tcoh = 1 \; \textrm{day}$, places the 
outliers within the effective region of the follow-up procedure.

Most wide parameter-space searches currently operate at $\Tcoh \sim \mathcal{O}(\textrm{hours})$. 
As demonstrated in \cite{PhysRevLett.124.191102,Abbott:2020mev}, CW candidates with uncertainties at such 
short coherence times can be successfully recovered by an MCMC follow-up at 
\mbox{$\Tcoh = 0.5\; \textrm{days}$}. 

We construct a hierarchical follow-up by imposing a first stage using \mbox{$\Tcoh = 0.5\; \textrm{days}$}
followed by a second stage using \mbox{$\Tcoh = 1\; \textrm{day}$}. Further stages are constructed by means
of Eq.~\eqref{eq:ladder_new} using $\N^{*} = 10^{4}$. The resulting coherence-time ladder, which is independent
of the parameter-space region and the prior specification due to the locality of the analysis, is collected in 
Table \ref{table:ladder}. As per the previous discussion on the sensitivity of the considered searches,
this ladder can be seamlessly applied to every one of the outliers under analysis.

\begin{table}
    \begin{tabular}{rccccc}
        \toprule
        Stage & 0 & 1 & 2 & 3 & 4\\
        \midrule
        $\Nseg$ & 500 & 250 & 55 & 5 & 1 \\
        $\Tcoh$ [days] & 0.5 & 1 & 5 & 55 & 270 \\
        \bottomrule
    \end{tabular}
    \caption{Coherence-time ladder constructed using \mbox{$\N^{*}=10^4$} and including an initial stage of 
    \mbox{$\Tcoh=0.5\;\textrm{days}$} before imposing \mbox{$\Tcoh=1 \;\textrm{days}$} and applying 
    the SuperSky metric. The results are independent of the parameter-space region at which the
    SuperSky metric was evaluated.}
    \label{table:ladder}
\end{table}

Table \ref{table:hyperparameters} specifies the hyperparameter setup of every MCMC stage, following the
setups employed in \cite{PhysRevLett.124.191102, Abbott:2020mev}. As demonstrated in 
Sec.~\ref{subsec:gaussian_noise}, this setup suffices to successfully follow up CW candidates within the
probed sensitivity range.

\begin{table}
    \begin{tabular}{ll}
        \toprule
        Hyperparameter & Value \\
        \midrule
        Parallel chains & 3 \\
        Walkers per chain & 100 \\
        Burn-in \& Production steps & 250 + 250\\
        \bottomrule
    \end{tabular}
    \caption{MCMC hyperparameter choices for each stage of the follow-up. The number of parallel chains
    equals the number of temperatures at which the likelihood is being sampled, following the recommendations
    in \cite{2016MNRAS.455.1919V, Ashton:2018ure}.}
    \label{table:hyperparameters}
\end{table}

The choice of initial priors is directly related to the outlier's uncertainty returned by each of the
analysis pipelines. Pipelines like \falcon{} or \eah{} return a well-determined parameter-space region in which
the outlier was found. The \viterbi{} pipelines, on the other hand, return only the frequency-evolution track of
each candidate, which can then be related to a certain parameter-space region if the stochastic contributions are
sub-dominant. The scope of a search also affects the prior
setup, as searches directed towards a particular sky position (such as the ones performed using \viterbi{}) 
allow us to place a narrower prior on the sky position of the outlier.
It is recommended in \cite{Ashton:2018ure} to choose a flat prior with fixed bounds containing the 
outlier's parameters. Instead, we use a set of Gaussian priors centered at the outlier's parameters
with scale parameters corresponding to the uncertainty in each dimension. After each MCMC step, we re-center
the priors on the median value of the resulting posterior distribution, taking half of the (centered) 
90\% credible interval as the new scale parameter, and re-sample the initial state of the MCMC ensemble. 
This particular setup ensures a fresh \mbox{start-up} at each stage of the ladder, preventing spurious samples 
dissociated from the ensemble to pollute the final results. Moreover, the use of unbounded priors prevents the 
follow-up from missing the true parameters of an outlier due to the presence of parameter-space correlations
\cite{Prix:2005gx, Prix:2006wm}.

The uncertainty associated to \falcon{} outliers is specified in \cite{Dergachev:2020fli} as
\begin{equation}
    \begin{gathered}
        \delta f_{0} = 5 \times 10^{-5}\;\textrm{Hz}\;,\\
        \delta f_{1} = 1 \times 10^{-12}\;\textrm{Hz/s}\;,\\
        \delta \theta = 0.06 \; \textrm{Hz} / f_{0}\;\textrm{rad}\;,
    \end{gathered}
    \label{eq:falcon_uncert}
\end{equation}
where $\delta \theta$ refers to the sky position of an outlier projected onto the ecliptic plane.
These uncertainties are conservatively lower than the canonical parameter-space resolution defined
in \cite{PhysRevD.70.082001} for a coherence time of \mbox{$\Tcoh=0.5\;\textrm{days}$}, meaning their 
corresponding parameter-space size is within acceptable values to ensure an effective MCMC stage
\cite{Abbott:2020mev, PhysRevLett.124.191102}. The \eah{} search reports uncertainties corresponding
to a coherence time of several months; since we start our follow-up at a lower coherence time, we used 
the same set of uncertainties as for the \falcon{} follow-up Eq.~\eqref{eq:falcon_uncert}.
\viterbi{} outliers were not reported as a parameter-space point, but as a
frequency band on which a significant frequency-evolution track was found; since both searches were targeted
at a particular sky position, we reduced the sky position uncertainty and increased the frequency uncertainty
by the same factor in order to cover all possible frequencies at a similar parameter-space size.

\section{Results}
\label{sec:results}
Before presenting results on the O2 outliers in Sec.~\ref{subsec:fu},
here we first describe an injection campaign in simulated Gaussian noise
to demonstrate the efficacy of the follow-up procedure
and calibrate a threshold on the newly introduced Bayes factor.

\subsection{Injections in Gaussian noise}
\label{subsec:gaussian_noise}

We characterize the behavior of $\logB$ using three sets of 100 artificial signals
at different signal strengths.
These are injected into Gaussian noise data compatible with the O2 observing run characteristics,
i.e. simulating data for both Advanced LIGO detectors and with
a duration of \mbox{$\Tobs=9\; \textrm{months}$}, using \texttt{lalapps\_Makefakedata\_v5}
\cite{lalsuite}. The actual O2 data stream covers 60\% of the duration of the run $\Tobs$
due to down time in the detectors (actual fractions are 65.3\% and 61.8\% for the H1 
and L1, respectively) \cite{O2Data}. Since SNR scales as the square root of observing time,
this would reduce the actual SNR of a signal \updated{to a fraction of 77\%.}
For the simulated Gaussian noise, we set the average amplitude spectral density to a fiducial
value of \mbox{$\sqrt{S_{\textrm{n}}} = 10^{-23} \;\textrm{Hz}^{-1/2}$}.
We injected the artificial signals at a fiducial frequency of $100\;\textrm{Hz}$, uniformly
spread across the whole sky and log-uniformly distributed in spindown parameter $f_{1}$ within
\mbox{$[-10^{-8}, -10^{-11}]\;\textrm{Hz/s}$}. The particular choice of a frequency band does
not affect the results of this analysis, since its effects are automatically taken into account
by parameter-space resolutions.

The CW amplitude $h_0$ is fixed in terms of the sensitivity depth \cite{PhysRevD.91.064007, Dreissigacker:2018afk}
\begin{equation}
    \depth = \frac{\sqrt{S_{\textrm{n}}}}{h_0}\;.
\end{equation}
Additionally, we define an \emph{effective} sensitivity depth by
explicitly including the effects of the cosine of the inclination angle $\iota$ \cite{Whelan:2015dha}:
\begin{equation}
    \deff = \frac{\depth}{\sqrt{\cos^4{\iota} + 6 \cos^2{\iota} + 1}}\;.
\end{equation}
We selected three depth values, enumerated in Table~\ref{table:efficiency}, bracketing the estimated 95\%
efficiency depth of the analyzed pipelines. The rest of the amplitude parameters were randomly drawn from 
uniform distributions \cite{Walsh:2016hyc}.

\begin{table}
    \begin{tabular}{cc}
        \toprule
        Depth $\left[ \textrm{Hz}^{-1/2}\right]$ & Efficiency (\%)\\
        \midrule
        40 & $97 \pm 2$\\
        60 & $98 \pm 1$\\
        80 & $96 \pm 2$\\
        \midrule
        Overall & $97 \pm 1$\\
        \bottomrule
    \end{tabular}
    \caption{Detection efficiencies for each set of 100 injections. An injection was labeled as detected
    if the final-stage posterior probability contained the injection parameters in its support. Error bars
    correspond to binomial errors.}
    \label{table:efficiency}
\end{table}

We start by estimating the detection efficiency of the follow-up. To do so, we run the full hierarchical
follow-up as specified in the previous section and count an injection as ``detected'' if the injection
parameters are within the final-stage posterior probability support. This criterion ensures the CW signals
are strong enough to guide the MCMC ensemble towards the relevant parameter space region, preventing
a signal from being lost. Results are reported in Table~\ref{table:efficiency}.
As expected from previous analyses in \cite{Ashton:2018ure}, we obtain a detection efficiency above
$95\%$ across the sensitivity range, meaning the follow-up is a suitable tool to further analyze
the selected set of outliers.

\begin{figure}
    \includegraphics[width=\columnwidth]{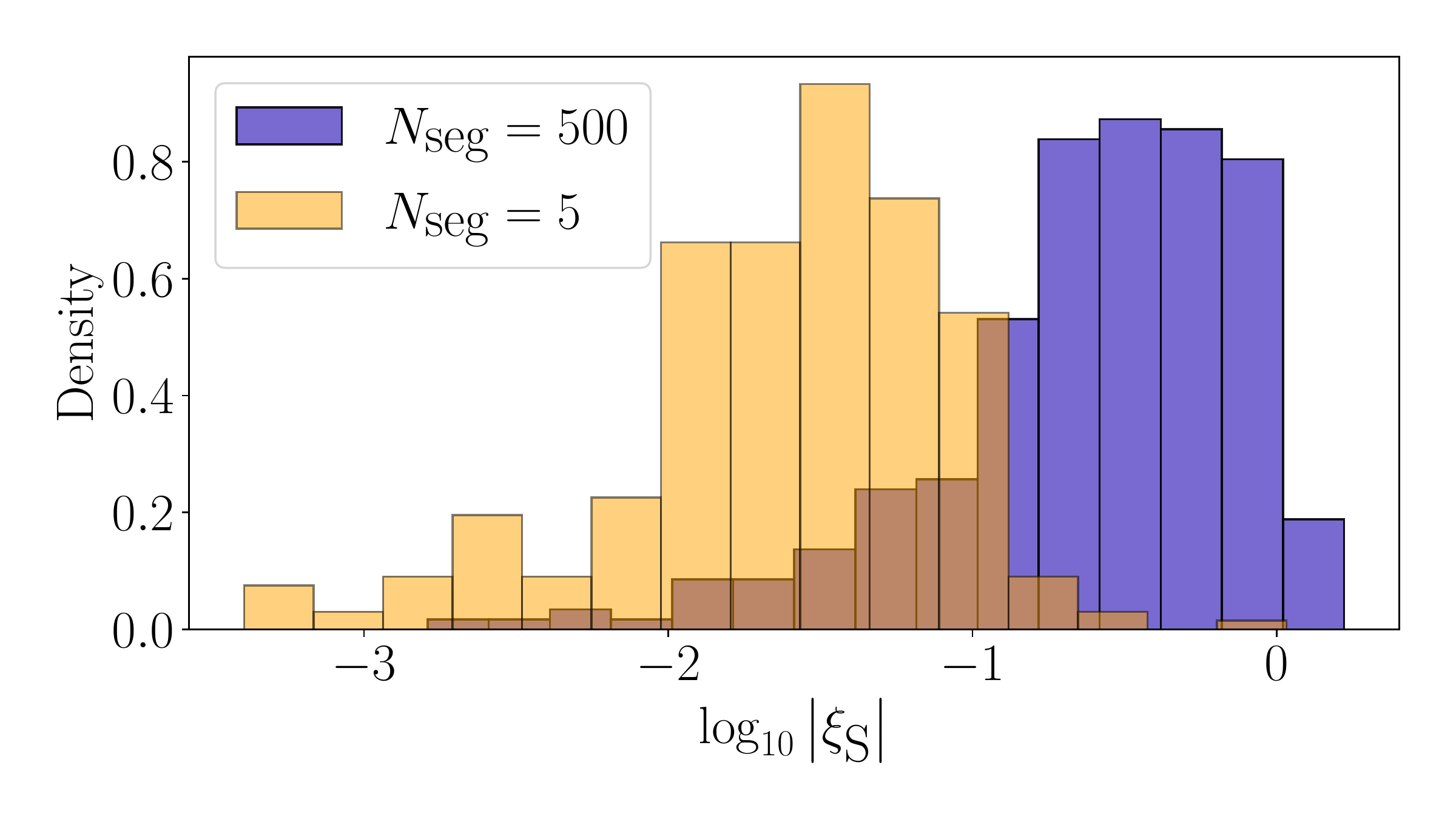}
    \caption{Distribution of $\abs{\x{S}}$ for the complete set of detected injections using different 
    semicoherent stages, namely \mbox{$\Nseg=500$} and \mbox{$\Nseg=5$}, as the reference to compute 
    $\mean{S}$ and $\std{S}$.}
    \label{fig:xi_S_histogram}
\end{figure}
\begin{figure}
    \includegraphics[width=\columnwidth]{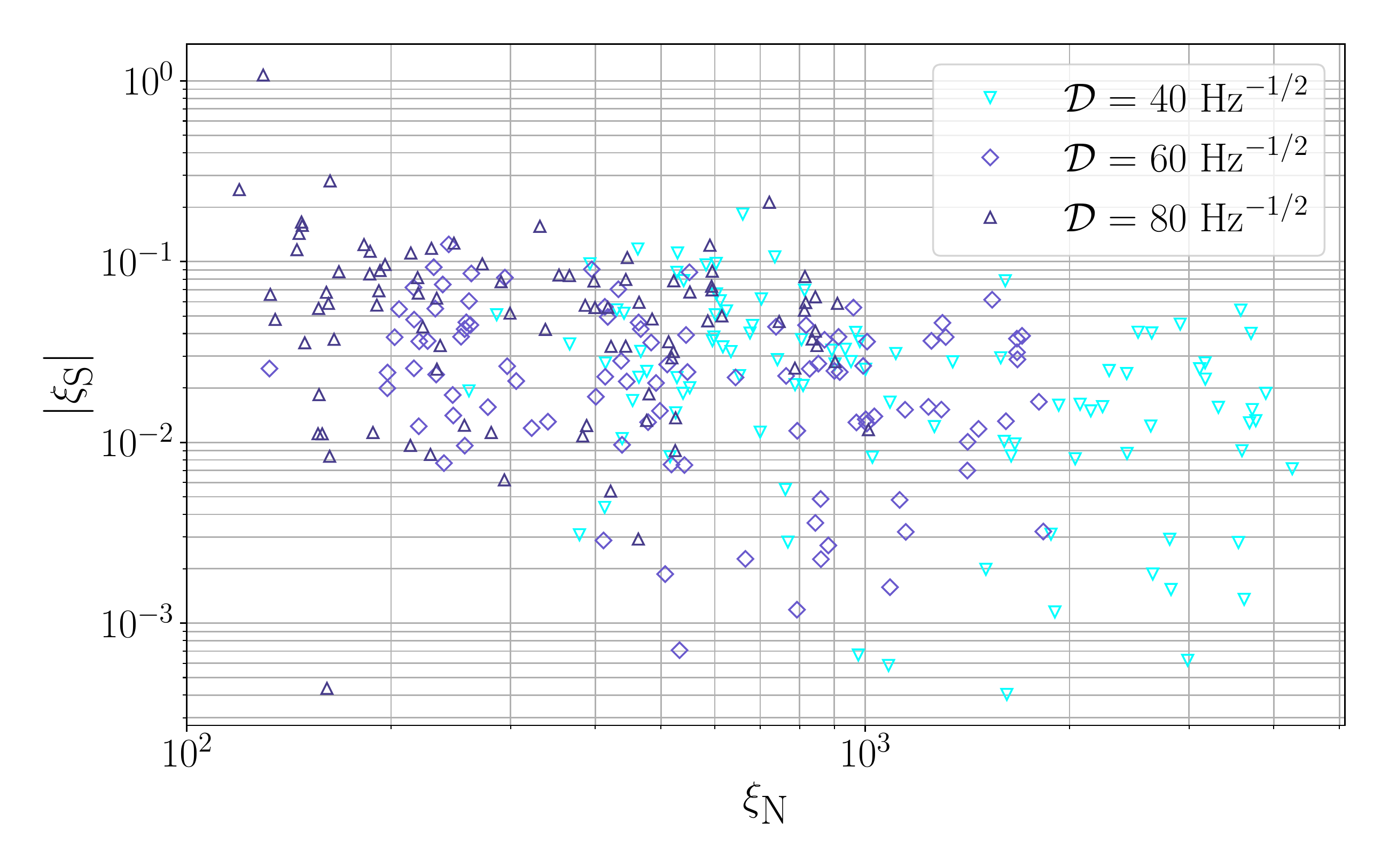}
    \caption{$(\x{S}, \x{N})$ plane for the complete set of detected injections using $\Nseg=5$ as the reference
    stage to compute $\mean{S}$ and $\std{S}$. The horizontal axis represents the discrepancy with respect to 
    the noise hypothesis, while the vertical axis represents the discrepancy with respect to the signal 
    hypothesis.}
    \label{fig:gaussian_noise_injections_S_vs_N}
\end{figure}

The computation of $\logB$ requires a particular semicoherent step from the ladder to be selected as the one
from which the expected fully-coherent distribution will be propagated.
As discussed in Sec.~\ref{subsec:signal_hypothesis}, using longer coherence times imposes a more restrictive
signal model, reducing the number of outliers due to the presence of detector artifacts and increasing the
significance of signal candidates (see e.g. Figure 6 of \cite{Ashton:2018ure}).
Fig.~\ref{fig:xi_S_histogram} shows the obtained distribution of signal-hypothesis discrepancies $\abs{\x{S}}$
for the complete set of detected injections with respect to two different stages. The use of a lower number
of segments (i.e. a longer coherence time) yields a tighter consistency with respect to the expected
distribution. We decide to carry out the analysis by taking the second-to-last stage of the ladder
($\Nseg=5$) as the reference from which the expected fully-coherent $\F$-statistic distribution will be
computed.

\begin{figure}
    \includegraphics[width=\columnwidth]{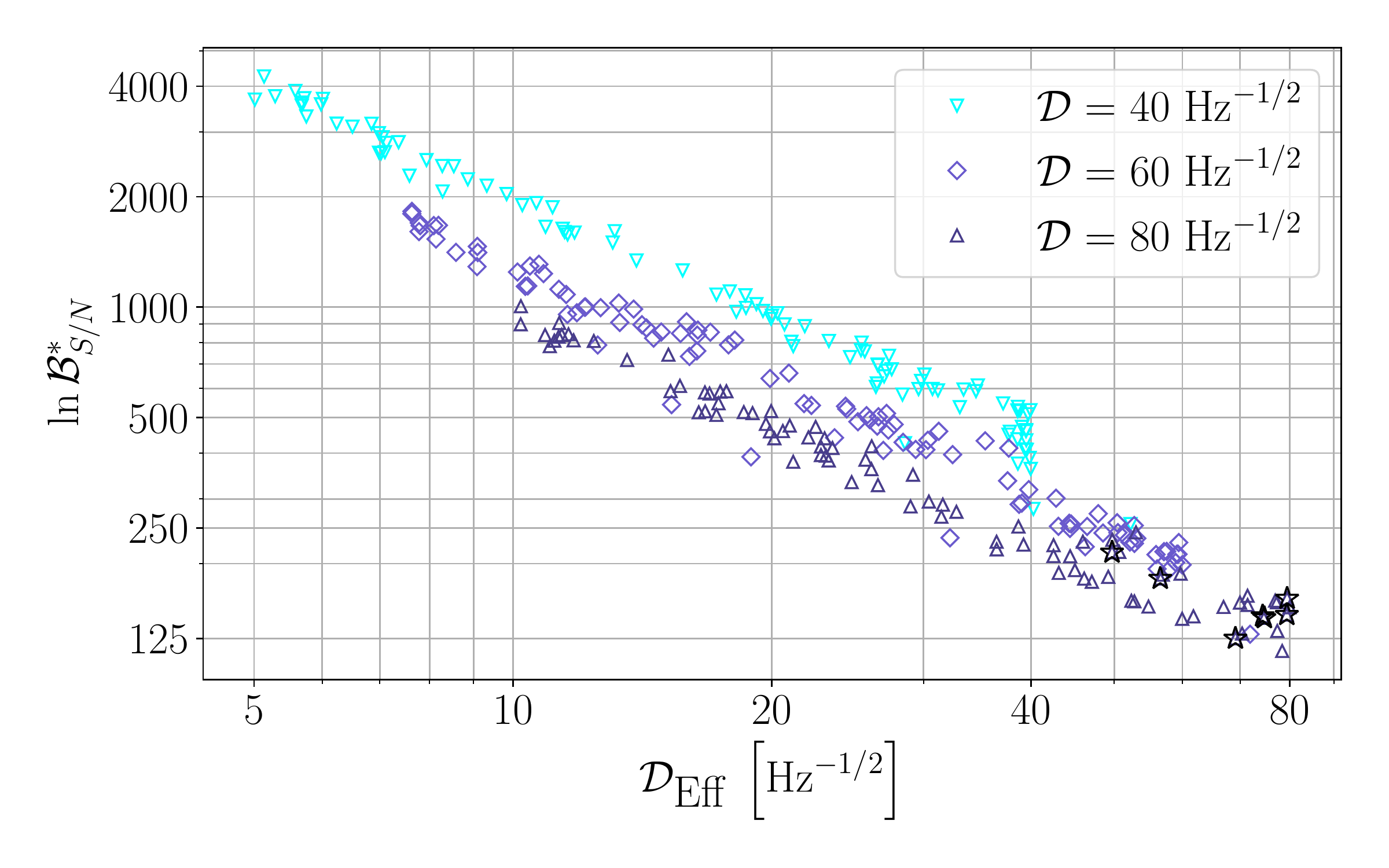}
    \caption{$\logB$ computed by applying the multi-stage MCMC follow up on the three sets of software injections. 
    Reference values were computed with respect to the $\Nseg=5$ stage and the Gaussian
    approximation was used to compute the signal contribution. Outliers marked by a star did not display an 
    ensemble-level volume shrinkage, as explained in the text.}
    \label{fig:gaussian_noise_injections_bayes}
\end{figure}

Figure \ref{fig:gaussian_noise_injections_S_vs_N} displays the distribution of injection results on the
$(\x{N}, \x{S})$ plane, showing the discrepancy of an outlier with respect to the noise and signal
hypotheses, respectively. The configuration is such that \mbox{$\x{S} \sim 0$} and
\mbox{$\x{N} \gg \x{S}$}, corresponding to case b) in Sec.~\ref{sec:b_star_sn}.
This means that the computation of the signal contribution to $\logB$ can be assumed to follow a
Gaussian distribution and, correspondingly,
Eq.~\eqref{eq:Bf} applies.
Figure \ref{fig:gaussian_noise_injections_bayes} shows the Bayes factor $\logB$ computed by comparing the last
two stages of the semicoherent ladder. The observed behavior $\logB \propto \deff^{-1}$ can be simply explained
by noting that $\deff$ is inversely proportional to SNR by definition \cite{Dreissigacker:2018afk}. 
\begin{figure}
    \includegraphics[width=\columnwidth]{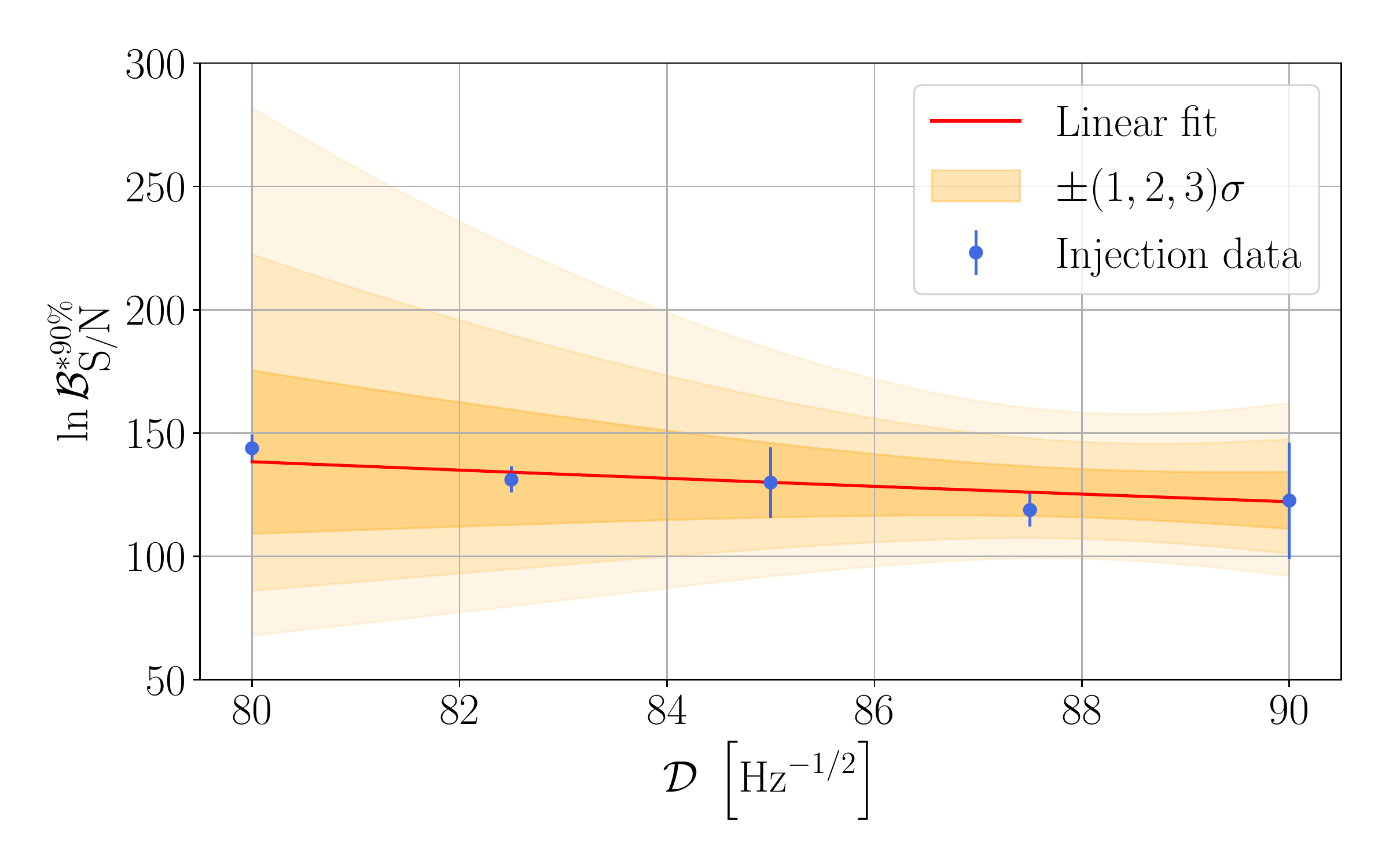}
    \caption{\updated{Estimation of 90\% detection-probability threshold on
    $\logB$ for different sensitivity depths beyond the injections shown in
    Fig.~\ref{fig:gaussian_noise_injections_bayes}. Each dot represents an empirical
    estimate of the 90\% detection-probability threshold using 100 simulated
    signals at a fixed depth value. Error bars correspond to the bootstrap
    standard deviation using 200 resamples of 50 samples each. The solid
    line and the associated envelopes represent a linear fit using
    \texttt{scipy.optimize.curve\_fit} \cite{2020SciPy-NMeth} with 1, 2, and 3 sigma
    uncertainties.}}
    \label{fig:90_threshold}
\end{figure}

\updated{This injection campaign covers the sensitivity ranges reported in Table~\ref{table:estimated_depth}
for all searches except one. The \eah{} search differs in that it was built as a subthreshold search: 
the reported outlier was thoroughly scrutinized using a variety of tools, 
including a fully-coherent analysis on O2 data, following a similar scheme as the one presented in this work. 
in order to assess the follow-up capabilities of our proposed method, we perform a second injection campaign 
akin to the previous one, covering the deepest \eah{} search sensitivity range.
Results are reported as 90\% detection-probability thresholds
in Fig.~\ref{fig:90_threshold}, following the approach proposed in~\cite{Dreissigacker:2018afk}. 
Were any of the considered outliers due to a genuine CW signal, the corresponding $\logB$ should lie
within the shaded region or higher. Based on this argument, we set a safe decision threshold at 
$\logB = 30$, also accounting for the reduced SNR in the real dataset due to detector downtime.
}

\begin{figure}
    \includegraphics[width=\columnwidth]{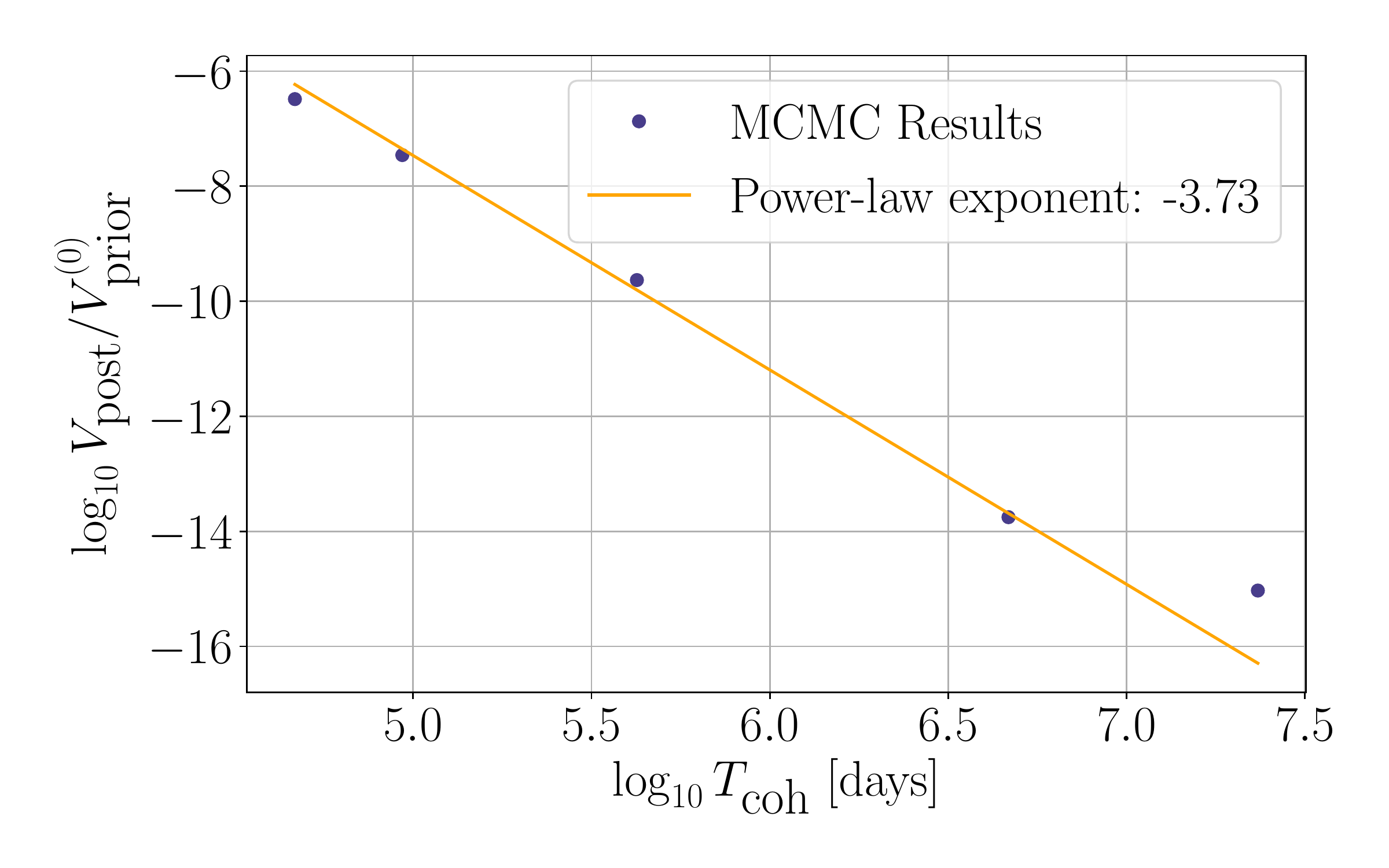}
    \caption{Posterior volume shrinkage of a successfully detected software injection. Parameter-space volumes are
    estimated by taking the product of parameter-wise 90\% central credible regions as explained in the text. 
    The vertical axis represents the posterior volume as a fraction of the initial prior volume. The slope of the
    log-log plot is an approximation to the inverse of the volume shrinkage $1/\updated{\gamma^{(j+1)}}$, where 
    the volume shrinkage $\updated{\gamma^{(j+1)}}$ was defined in Eq.~\eqref{eq:v_jp1}.}
    \label{fig:volume_shrinkage_example}
\end{figure}

\begin{figure}
    \includegraphics[width=\columnwidth]{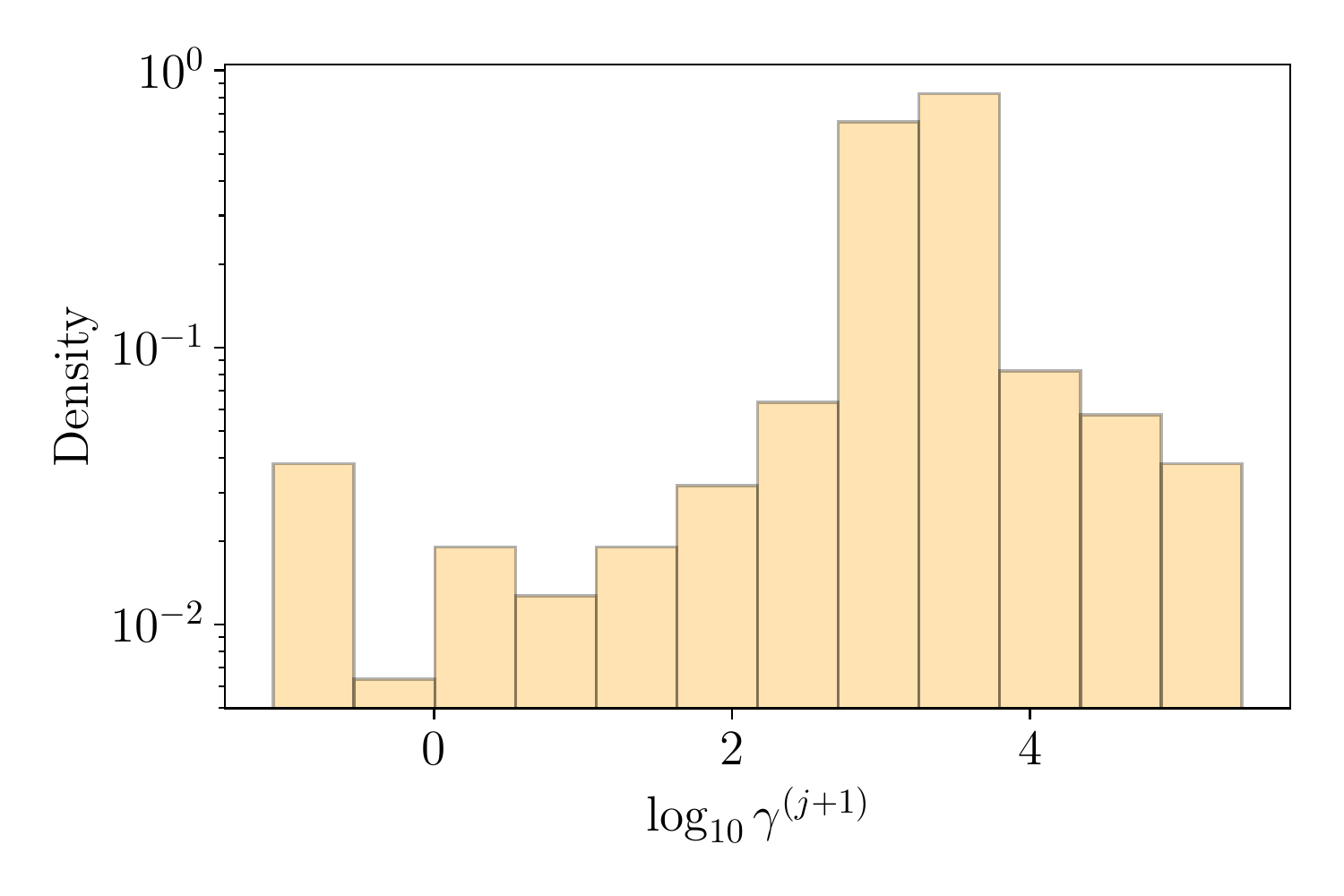}
    \caption{Distribution of ($\log_{10}$) posterior volume shrinkage rates of the detected injections. The
    prominent peak is within the order of magnitude of the expected value according to Eq.~\eqref{eq:v_jp1}.
    The presence of \emph{negative} shrinkage rates for a few injections is discussed in the text.}
    \label{fig:log_v_distribution}
\end{figure}

Lastly, we comment on the behavior of the multi-stage MCMC itself in terms of the volume shrinkage rate
introduced in Sec.~\ref{subsec:ladder}.
Figure \ref{fig:volume_shrinkage_example} shows the behavior of the posterior
volume of a successfully detected injection. The quantities $\textrm{V}^{(0)}_{\textrm{prior}}$ and 
$\textrm{V}_{\textrm{post}}$ represent approximations to the initial prior volume at the first stage of the
ladder and the posterior volume after each of the MCMC stages. These quantities are computed by taking the product
of parameter-wise central 90\% credible intervals, since we are only interested in the overall scaling along the
coherence-time ladder. The volume shrinkage shows a power-law behavior, the exponent of which (i.e. the slope in
log-log scale) should be approximately given by \mbox{$\log_{10}\updated{\gamma}^{(j+1)}\sim 4$} 
from Eq.~\eqref{eq:v_jp1}.
The same procedure is performed on the complete set of detected injections, collecting the power-law indices into
a histogram in Fig.~\ref{fig:log_v_distribution}. The rate of volume shrinkage accumulates a prominent peak within 
the order of magnitude of the expected result.

Figure \ref{fig:log_v_distribution} also displays a small set of injections for which the MCMC ensemble
did not produce a clear shrinkage of the (approximated) central 90\% credible region, even though the
true injection parameters are contained within said region. Their corresponding $\logB$ values are marked
using stars in Fig.~\ref{fig:gaussian_noise_injections_bayes}, belonging to the weakest set of performed
injections. \updated{This is a consequence of the parameter-space structure in the vicinity of a
signal~\cite{Prix:2006wm}: The effective (squared) SNR recovered by a template falls off as a
linear function of the mismatch with respect to the true signal parameters. Strong injections, associated
to higher SNR values, are able to sustain an \mbox{$\F$-statistic} above background throughout a wider
parameter-space region than weak injections. Weak injections, as a result, require tighter priors to
display a similar behavior to that of stronger injections.} The fact that the ensemble is unable to focus
into a particular parameter-space region, however, is still compatible with a good recovery of $\logB$,
as for that it is only required to \emph{sample} the region of interest during the production stage.
This is in fact the principle upon which the application of a single-stage MCMC follow-up as a simple
veto was based in~\cite{PhysRevLett.124.191102, Abbott:2020mev}, and can be justified by interpreting the
MCMC follow-up as being equivalent to a search starting from a \emph{random} template bank at higher
mismatches than traditionally suggested in CW searches \cite{Messenger:2008ta, Allen:2021eju}.

\subsection{Follow-up of CW outliers from Advanced LIGO O2 data}
\label{subsec:fu}

We now report the results of the multi-stage MCMC-based follow-up
on the set of outliers described in Sec.~\ref{subsec:o2_outliers}
in terms of the obtained $(\x{N}, \x{S})$ values and the corresponding $\logB$.

\begin{figure}
    \includegraphics[width=\columnwidth]{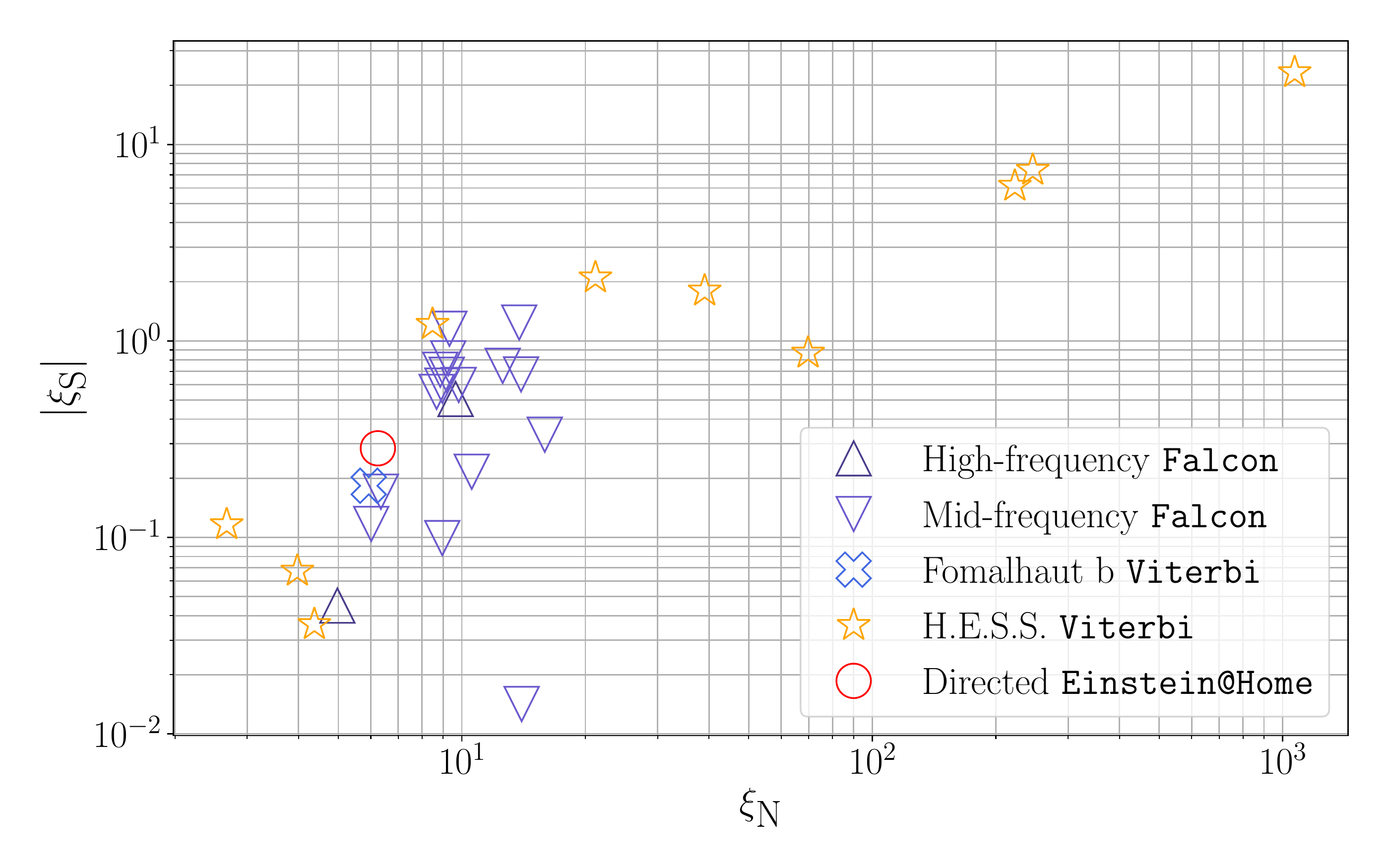}
    \caption{$(\x{N}, \x{S})$ plane associated to the outliers found in O2 data by the specified searches.}
    \label{fig:outliers_S_N}
\end{figure}

Figure \ref{fig:outliers_S_N} shows each pipeline's outliers across the $(\x{N}, \x{S})$ plane, quantifying 
their discrepancy with respect to the noise and signal hypotheses. 
\updated{
The bulk of outliers show discrepancies with respect to the signal hypothesis, quantified by $\abs{\x{S}}$, 
an order of magnitude larger than those displayed by software injections 
in Fig.~\ref{fig:gaussian_noise_injections_S_vs_N}.
Discrepancies with respect to background noise, quantified by $\x{N}$, are more than an order of magnitude lower.
}
The retrieved values of $\mean{N}$ and $\std{N}$ are within the brackets obtained in Gaussian noise, 
suggesting these results are not because of an elevated background noise but rather a low SNR associated to the outliers.
We note the presence of three \viterbi{} outliers at high values of $\x{N}$, namely J1831-0952@19.9991Hz,
J1849-0001@26.3410Hz and J1831-0952@15.4012Hz; and another marginal pair stemming from the same pipeline in 
the middle ground, namely J1718-3825@17.5034Hz and J1831-0952@14.5018Hz.

\begin{figure}
    \includegraphics[width=\columnwidth]{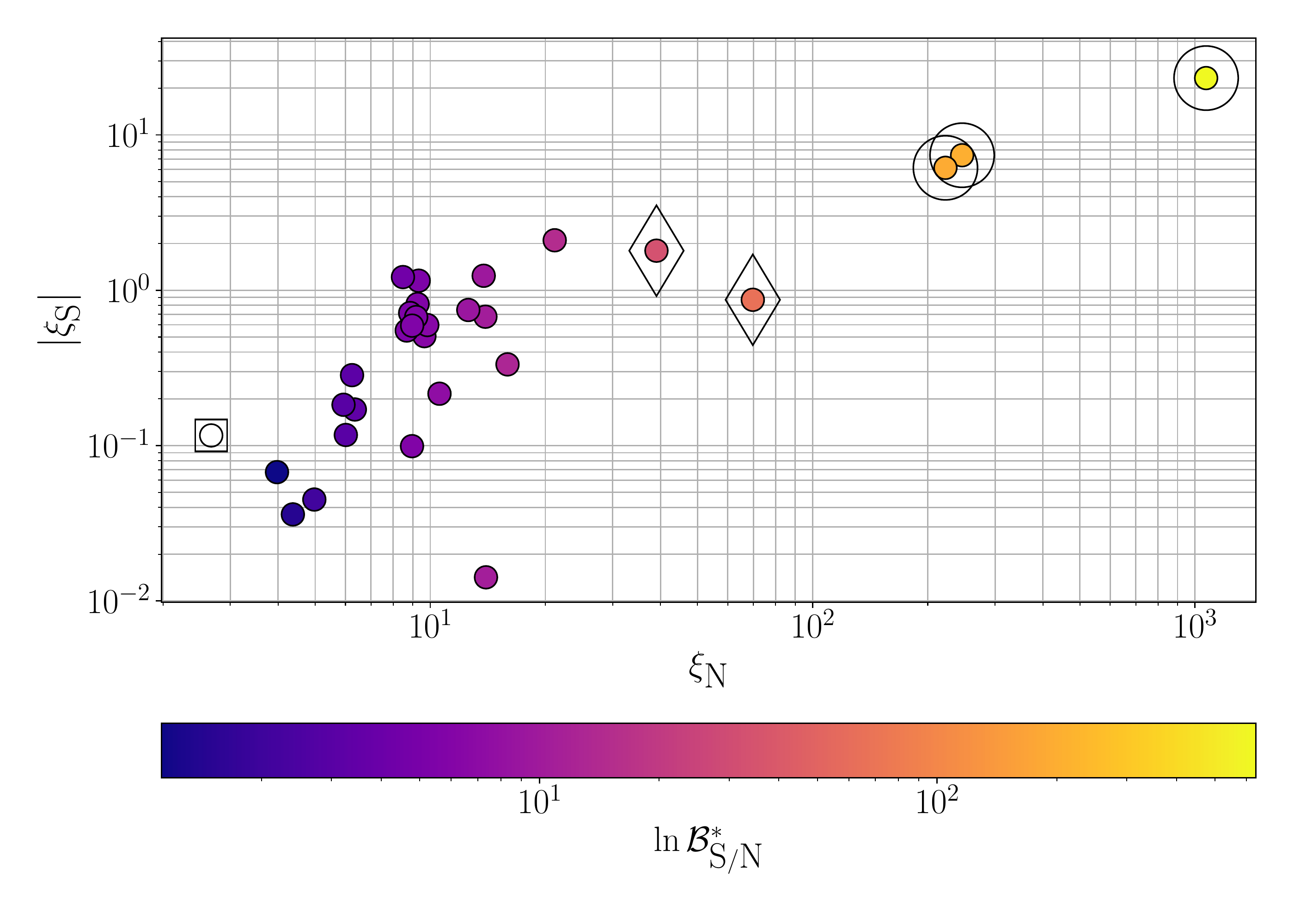}
    \caption{$\logB$ values obtained after the hierarchical MCMC follow-up of O2 outliers.
    Relative outlier positions in this figure are consistent with Fig.~\ref{fig:outliers_S_N}.
    Outliers enclosed by circles and diamonds score a $\logB$ value above 30. The outlier enclosed by 
    a square returns a negative value of $\logB$ and is displayed as white due to the logarithmic color scale.}
    \label{fig:outliers_S_N_BstarSN}
\end{figure}

We compute $\logB$ by numerically integrating Eq.~\eqref{eq:fc_from_semi} due to the regime in which outliers
are placed. Results are listed in Table \ref{table:loudest} and displayed in Fig.~\ref{fig:outliers_S_N_BstarSN}.
\updated{Five outliers score over the decision threshold \mbox{$\logB = 30$}, all of them related to the H.E.S.S.}
\viterbi{} pipeline.

\updated{The first set of outliers, J1831-0952@15.4012Hz, J1831-0952@19.9991Hz, and J1849-0001@26.3410Hz,
is highlighted using circular markers in Fig.~\ref{fig:outliers_S_N_BstarSN}.}
The original search \cite{Beniwal:2021hvc} ascribed them to instrumental artifacts in the L1 detector. 
We confirm that to be the case for the outlier \mbox{J1831-0952@19.9991Hz}:
the loudest fully-coherent \mbox{$\F$-statistic} recovered by our follow-up is located at 
\mbox{$f_{0} \simeq 20.0011\;\textrm{Hz}$}, crossing a well-known instrumental comb at 
both LIGO detectors \cite{2018PhRvD..97h2002C}. For outlier \mbox{J1849-0001@26.3410Hz}, we note the 
presence of a hardware injection
(a CW-like signal simulated by direct actuation of the interferometer mirrors,
used to test calibration and analysis pipelines)
at \mbox{$f_{0} \simeq 26.3396 \;\textrm{Hz}$} with an amplitude corresponding
to \mbox{$\depth \sim \mathcal{O}(1\;\textrm{Hz}^{-1/2})$} \cite{Biwer:2016oyg, O2Datadoi}.
Even though the spindown and sky positions are completely mismatched, such strong artificial signals are 
known to produce loud candidates across wide parameter space regions 
\cite{Aasi:2013jya, Papa:2016cwb, Abbott:2017pqa, Piccinni:2019zub, PhysRevLett.124.191102}.
We are unable to relate \mbox{J1831-0952@15.4012Hz} to any of the listed narrow spectral artifacts in 
\cite{2018PhRvD..97h2002C,O2Datadoi}.

\begin{figure}
    \includegraphics[width=\columnwidth]{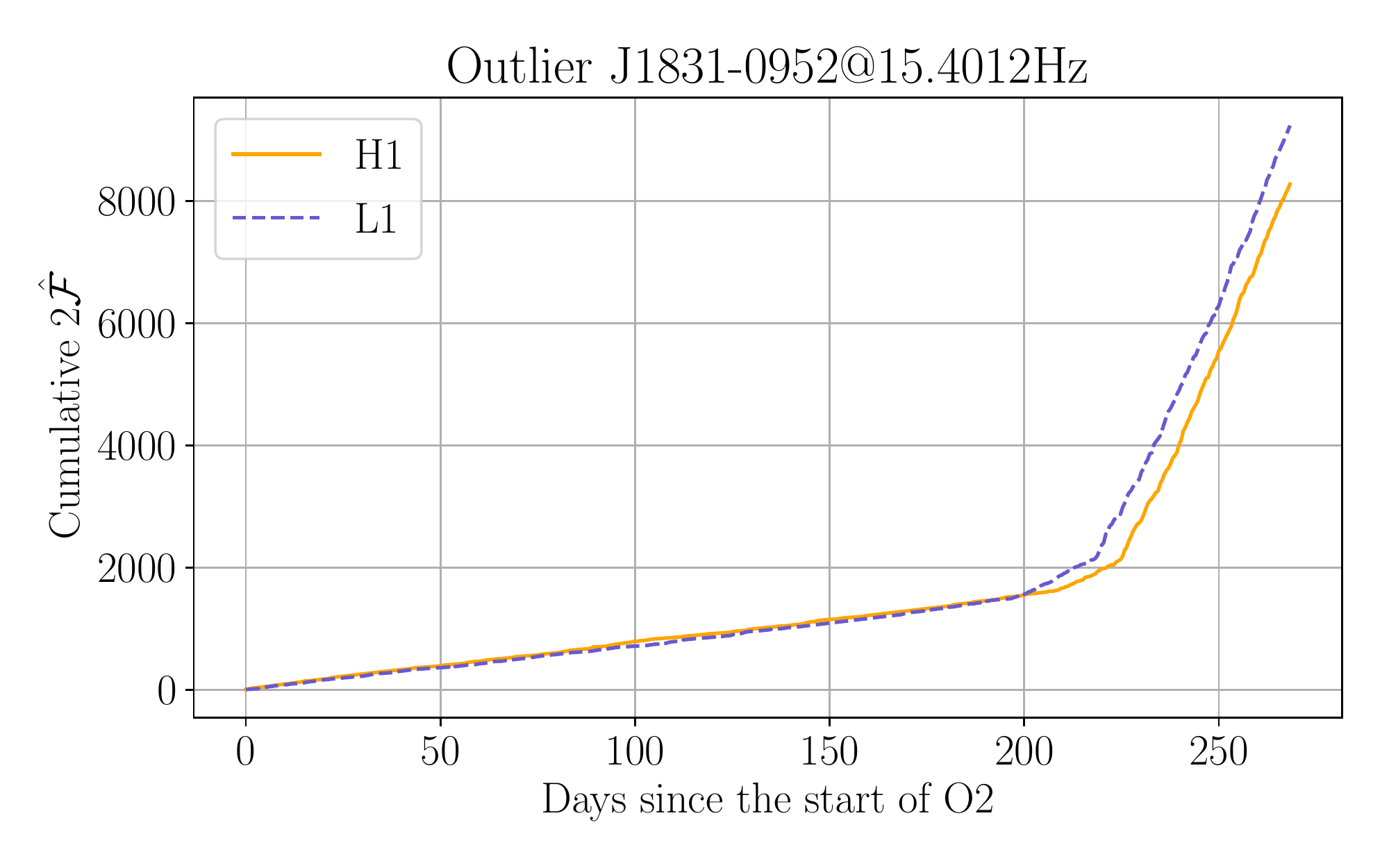}
    \includegraphics[width=\columnwidth]{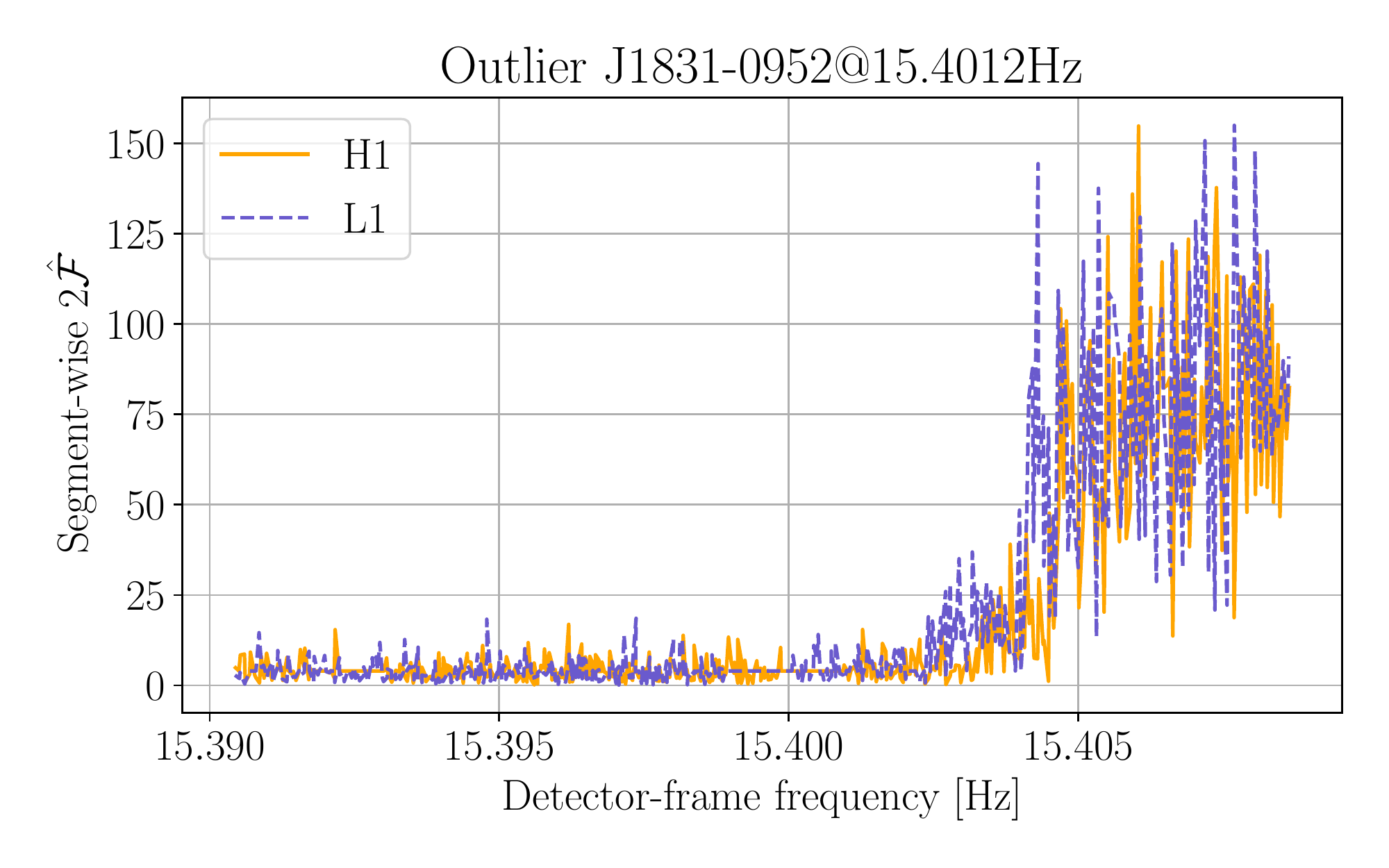}
    \caption{\updated{Segment-wise $2\hat\F$ accumulation of the loudest template associated to the 
    outlier \mbox{J1831-0952@15.4012Hz} throughout the observing run using 500 coherent segments. 
    The solid orange line shows the results using the LIGO Hanford detector (H1) only, while the dashed blue 
    line shows the results using the LIGO Livingston detector (L1). The upper panel shows the $2\hat\F$ 
    accumulation throughout the duration  of the run. The lower panel shows the \mbox{segment-wise} $2\hat\F$ 
    values per frequency bin. Frequency values are computed evaluating the frequency-evolution template at 
    the starting time of each segment.}}
    \label{fig:viterbi_outliers}
\end{figure}
A manual check of the segment-wise semicoherent \mbox{$\F$-statistic} values of \mbox{J1831-0952@15.4012Hz} 
reveals a rapid accumulation of $\F$-statistic as the frequency evolution crosses a narrow sub-band. This kind of 
behavior, shown in Fig.~\ref{fig:viterbi_outliers}, is inconsistent with a CW signal and usually can be 
related to instrumental artifacts, but said identification becomes more difficult at low frequencies as 
they are populated by a wider variety of noise sources. As a result, outlier \mbox{J1831-0952@15.4012Hz} 
is also likely related to an instrumental artifact.

The second group of outliers is enclosed by diamonds in Fig.~\ref{fig:outliers_S_N_BstarSN}.
\updated{Outlier \mbox{J1831-0952@14.5018Hz}'s loudest candidate is recovered at 
\mbox{$f_{0} \simeq 14.4953 \;\textrm{Hz}$}. This is consistent with a $1\;\textrm{Hz}$ comb with an offset of 
$0.5\;\textrm{Hz}$.}\footnote{\updated{The spindown value reported by the original search 
is such that also positive values are covered by the initial prior volume, and indeed our followup recovered 
the loudest candidate at positive spindown.}}
Another harmonic of the same comb can be related to outlier \mbox{J1718-3825@17.5034Hz}, whose loudest candidate 
is located at $f_{0} \simeq 17.5005 \;\textrm{Hz}$.

\updated{The remaining outliers from all searches return a $\logB$ value below the decision threshold
$\logB = 30$. For completeness, we list the parameters recovered by the final follow-up stage in 
Table \ref{table:loudest}. We highlight outlier J1713, initially found by the \eah{} search and scoring below
our decision threshold. This result is consistent with the latest Einstein@Home search
for J1713 reported in \cite{Ming:2021xtz}, covering up to 400\,Hz in O2 data,
 not finding any significant outliers.}

\begin{table*}
    \begin{tabular}{lrrrrrr}
\toprule
   Outlier ID &  $f_0$ [Hz] &  $f_1$ [Hz/s] &  $f_2$ [Hz/$\textrm{s}^2$] & $\alpha$ [rad] &  $\delta$ [rad] & $\logB$\\
\midrule
    Falcon 4 & 1891.756615 & $ -6 \cdot 10^{-12}$ & $1 \cdot 10^{-20}$ & 2.987285 & 1.005941 & 2.10 \\
    Falcon 5 & 1892.991046 & $ -2 \cdot 10^{-12}$ & $-1.8 \cdot 10^{-19}$ & 3.778973 & $-$0.816265  & 6.46 \\
    Falcon 15 & 900.218764 & $ $-$1 \cdot 10^{-12}$ & $7 \cdot 10^{-20}$ & 2.084412 & $-$0.102330 &  10.49 \\
    Falcon 19 & 514.148984 & $ 6 \cdot 10^{-12}$ & $3.7 \cdot 10^{-19}$ & 2.170669 & 0.092978 & 9.04 \\
    Falcon 23 & 1001.366278 &  $2 \cdot 10^{-12}$ & $4.9 \cdot 10^{-19}$ & 1.355553 & $-$0.769952 &  5.88 \\
    Falcon 24 &  676.195493 & $3\cdot 10^{-12}$ & $-2.6 \cdot 10^{-19}$ & 3.846438 & $-$0.102138 & 5.80 \\
    Falcon 25 &  744.219196 & $2 \cdot 10^{-12}$ & $0.4 \cdot 10^{-20}$ & 3.344781 & 0.612270 &  7.42 \\
    Falcon 29 &  512.490782 & $-8 \cdot 10^{-12}$ & $-2 \cdot 10^{-20}$ & 2.468688 & $-$0.041880 & 3.37 \\
    Falcon 31 &  983.151151 & $-1 \cdot 10^{-12}$ & $-7.0 \cdot 10^{-19}$ & 3.562362 & 0.018926 & 3.06 \\
    Falcon 34 &  886.880063 & $-2 \cdot 10^{-12}$ & $-2.1 \cdot 10^{-19}$ & 4.912748 & $-$0.703663 & 10.85 \\
    Falcon 35 &  988.373241 & $2 \cdot 10^{-12}$ & $-1.3 \cdot 10^{-19}$ &  0.982043 & 0.778393 & 6.48 \\
    Falcon 39 &  514.291753 & $3 \cdot 10^{-12}$ & $-3.5 \cdot 10^{-19}$ &  0.569150 & $-$0.128791 & 5.47 \\
    Falcon 40 &  831.988457 & $-3 \cdot 10^{-12}$ & $-1 \cdot 10^{-20}$ & 4.917884 & 1.160566 & 5.48 \\
    Falcon 41 & 873.524663 & $3 \cdot 10^{-12}$ & $4 \cdot 10^{-20}$ & 0.619107 & $-$0.189295 & 5.77 \\ 
    Falcon 42 &  895.421995 & $1 \cdot 10^{-12}$ & $-1.9 \cdot 10^{-19}$ & 5.105728 & 0.249030 & 5.37 \\
    Falcon 43 & 1224.745693 & $1 \cdot 10^{-12}$ & $-1.2 \cdot 10^{-19}$ & 1.715372 & 0.196097 & 5.70 \\
    Falcon 45 & 698.728033 & $1 \cdot 10^{-12}$ & $1.6 \cdot 10^{-19}$ & 4.557448 & $-$0.723930 & 12.69 \\
    Falcon 46 & 1095.557373 & $-4 \cdot 10^{-12}$ & $-5.7 \cdot 10^{-19}$ & 4.354405 & $-$0.260292 & 9.60 \\
      J1713 &  368.801590 & $-4.380 \cdot 10^{-9}$ & $1.18 \cdot 10^{-18}$ & 4.511570 & $-$0.694137 & 3.21\\
    Fomalhaut b &  876.517914 & $-4.2979 \cdot 10^{-10}$ & $-5.67 \cdot 10^{-18}$ & 6.011153 & 0.516952 & 2.96 \\
    J0534+2200  & 29.813469 & $-2.3430 \cdot 10^{-10}$ & $1.158 \cdot 10^{-17}$ & 1.461040 & 0.385286 & $-$0.16 \\
    J1420--6048  & 14.511112 & $-2.5 \cdot 10^{-11}$ & $1.364 \cdot 10^{-17}$ & 3.750570 & $-$1.061001 & 1.46 \\
    J1420--6048  & 19.512364 & $-4.4 \cdot 10^{-11}$ & $9.80 \cdot 10^{-18}$ & 3.753242 & $-$1.061227 & 1.12 \\ 
    J1420--6048  & 29.526774 & $3.9 \cdot 10^{-11}$ & $5.30 \cdot 10^{-18}$ & 3.753011 & $-$1.060915 & 4.46 \\
    J1718--3825  & 17.500500 & $-4.0 \cdot 10^{-11}$ & $7.58 \cdot 10^{-18}$ & 4.528719 & $-$0.670563 & \textit{33.45} \\
    J1831--0952  &  14.495361 & $2.95 \cdot 10^{-10}$ & $-3.89 \cdot 10^{-18}$ & 4.848071 & $-$0.172356 & \textit{65.64} \\ 
    J1831--0952  &  15.389002 & $8.72 \cdot 10^{-10}$ & $-2.980 \cdot 10^{-17}$ & 4.853052 & $-$0.165697 & \textbf{203.35} \\
    J1831--0952  & 20.0016854 & $-7.13 \cdot 10^{-10}$ & $7.017 \cdot 10^{-17}$ & 4.859380 & $-$0.179332 & \textbf{633.204} \\
    J1849--0001  & 26.3062476 &$ -6.4 \cdot 10^{-11}$ & $1.101 \cdot 10^{-17}$ & 4.850380 & $-$0.000331 & 14.71 \\
    J1849--0001  &   26.333433 & $-7.1 \cdot 10^{-11}$ & $2.297 \cdot 10^{-17}$ & 4.850122 & 0.003055 & \textbf{192.71} \\
\bottomrule
\end{tabular}
     \caption{Loudest template recovered by the multi-stage MCMC follow up for each of the analyzed outliers. 
    Boldface and italic $\logB$ values correspond to the two sets of outliers highlighted with circles and diamonds
    in Fig~\ref{fig:outliers_S_N_BstarSN}, respectively.} 
    \label{table:loudest} 
\end{table*}
 
\section{Conclusion}
\label{sec:conclusion}
We have introduced the first complete framework to analyze outliers from arbitrary CW searches using
a multi-stage MCMC-based follow-up. After demonstrating its general behavior on Gaussian noise,
we applied it to a set of 30 outliers obtained by different CW search pipelines on O2
Advanced LIGO data \cite{Dergachev:2020fli, Dergachev:2020upb, Jones:2020htx, 
Beniwal:2021hvc, Papa:2020vfz}.

The procedure constructs a Bayes factor comparing whether the behavior of the $\F$-statistic
across different stages of the analysis
is more consistent with the presence of a signal rather than with pure noise.
The expected evolution of this detection statistic as the follow-up progresses can be derived from first
principles. The noise contribution is described by applying extreme value theory to samples of background 
noise data. These samples can be obtained by sampling shifted sky positions with respect to the 
outliers, blinding the analysis from the presence of a signal.

\updated{
The application of a multi-stage MCMC follow-up deemed 25 of the analyzed outliers as
less consistent with a standard CW signal than with background noise.
The remaining five outliers passed the specified threshold and were manually inspected.
Four of them were successfully associated to known instrumental artifacts in the Advanced LIGO detectors.
The fifth outlier displays a behavior inconsistent with a CW signal but consistent with an instrumental
artifact; the exact instrumental cause, however, could not be identified.}

Although the outliers were analyzed assuming a standard signal model corresponding to an isolated CW source,
the framework presented here (and the \texttt{PyFstat} software used
\cite{Ashton:2018ure,Keitel2021,ashton_gregory_2021_4542822})
can be seamlessly applied to more general models,
such as sources in binary systems \cite{PhysRevLett.124.191102, Abbott:2020mev}, sources producing glitches
\cite{Ashton:2018qth}, and long gravitational-wave transient signals \cite{Prix:2011qv, Keitel:2018pxz}.

This represents the first application of a multi-stage MCMC-based follow-up to CW outliers 
from real data. The scalability of this development is such that it can be taken as a default follow-up 
strategy to outliers produced by virtually any CW search, as long as they can be related 
to a well-defined parameter space region. This allows for the general application of long-coherence 
follow-ups, massively reducing the complexity \updated{associated with the setup} and calibration of ad hoc vetoes
in CW searches.
 
\section*{Acknowledgements}
We gratefully thank Luana M. Modafferi, Luca Rei, and the LIGO-Virgo-KAGRA Continuous Wave working group
for many fruitful discussions. 
We thank an anonymous referee for valuable comments which enhanced the quality of this manuscript.
This work was supported by European Union FEDER funds, the Ministry of Science, 
Innovation and Universities and the Spanish Agencia Estatal de Investigaci\'on grants
PID2019-106416GB-I00/AEI/10.13039/501100011033,  RED2018-102661-T,    RED2018-102573-E,    Comunitat Aut\`onoma de les Illes Balears through Conselleria de Fons Europeus, Universitat i Cultura
and the Direcci\'o General de Pol\'itica Universitaria i Recerca with funds from the Tourist Stay Tax Law ITS 2017-006 (PRD2018/24),
Generalitat Valenciana (PROMETEO/2019/071),  
EU COST Actions CA18108, CA17137, CA16214, and CA16104.
R.~T.~is supported by the Spanish Ministry of Science, Innovation and Universities (ref.~FPU 18/00694).
D.~K.~is supported by the Spanish Ministry of Science, Innovation and Universities (ref.~BEAGAL 18/00148)
and cofinanced by the Universitat de les Illes Balears.
The authors thankfully acknowledge the computer resources at Picasso and the technical support provided by
Barcelona Supercomputing Center - Centro Nacional de Supercomputaci\'on through grants
No. AECT-2020-3-0022 and No. AECT-2021-1-0029 from the Red Espa\~nola de Supercomputaci\'on (RES).
 This research has made use of data, software and/or web tools obtained from the Gravitational Wave Open Science Center (https://www.gw-openscience.org/), a service of LIGO Laboratory, the LIGO Scientific Collaboration and the Virgo Collaboration. LIGO Laboratory and Advanced LIGO are funded by the United States National Science Foundation (NSF) as well as the Science and Technology Facilities Council (STFC) of the United Kingdom, the Max-Planck-Society (MPS), and the State of Niedersachsen/Germany for support of the construction of Advanced LIGO and construction and operation of the GEO600 detector. Additional support for Advanced LIGO was provided by the Australian Research Council. Virgo is funded, through the European Gravitational Observatory (EGO), by the French Centre National de Recherche Scientifique (CNRS), the Italian Istituto Nazionale della Fisica Nucleare (INFN) and the Dutch Nikhef, with contributions by institutions from Belgium, Germany, Greece, Hungary, Ireland, Japan, Monaco, Poland, Portugal, Spain.
  This paper has been assigned document number LIGO-P2100187.
 
\appendix
\section{On the distribution of the maximum $\F$-statistic and the effective number of templates}
\label{sec:appendix}
The validity of using an effective number of templates to fit Eq.~\eqref{eq:prob_max}
for the expected maximum $\F$-statistic from a search over a certain actual number of templates,
\updated{in the presence} of \mbox{\emph{non-independent}} templates, has been discussed in the CW literature
\cite{Wette:2011eu, Dreissigacker:2018afk}. We attempt to shed some light on the topic using extreme 
value theory. Concretely, we analyze the toy model posed in Appendix D of \cite{Dreissigacker:2018afk}.

The basic point in \cite{Dreissigacker:2018afk} is that the presence of correlated templates not only 
changes the effective number of templates, but also the ``functional form'' of the resulting distribution, 
rendering Eq.~\eqref{eq:prob_max} inaccurate. As an example, a toy model is constructed
by generating a time series of zero-mean unit-variance Gaussian noise and computing the power of 
its Fourier transform. By choosing a suitable normalization, said power is the squared sum of two
identical zero-mean Gaussian variables, following a chi-squared distribution with two
degrees of freedom. This distribution can be properly fitted using Eq.~\eqref{eq:prob_max}, and the
effective number of templates $\N'$ is consistent with the number of frequency samples 
\mbox{$\N = N/2 -1$}, where $N$ is the number of elements from the original time series.
Correlated templates are then introduced by over-resolving the Fourier transform applying zero-padding 
to the time series. The resulting distribution cannot be properly fitted using Eq.~\eqref{eq:prob_max}.
The effective number of independent templates $\N'$ is found to \emph{increase} with the length of 
zero-padding, but it remains bounded by the actual number of power samples $\N$.

We  provide an explanation for the two main issues raised in \cite{Dreissigacker:2018afk}, 
namely what is the actual ``functional form'' of the target distribution and why the effective
number of templates seems to increase as more correlated templates are included.

Let $x_{n=1, \dots, N}$ be a zero-mean unit-variance Gaussian process. We define its Fourier transform
as
\begin{equation}
   \tilde{x}_{k} = \sum_{n=0}^{N-1} x_{n} e^{-2 \pi i \, n \frac{k}{N}}
   \label{eq:fourier}
\end{equation}
where $k=0, \dots N-1$. Since $x_{n} \in \mathbb{R}$, the real and imaginary parts of Eq.~\eqref{eq:fourier}
follow a zero-mean Gaussian distribution
\begin{equation}
    \begin{split}
        \mathfrak{R} \tilde{x}_{k} \sim \mathrm{Gauss}(0, \sqrt{N/2})\\
        \mathfrak{I} \tilde{x}_{k} \sim \mathrm{Gauss}(0, \sqrt{N/2})\\
    \end{split}\;.
    \label{eq:fourier_stats}
\end{equation}
We then define power as
\begin{equation}
    \tilde\rho
    = \left(\sqrt{\frac{2}{N}}\:\mathfrak{R} \tilde{x}_{k}\right)^2  
    + \left(\sqrt{\frac{2}{N}}\:\mathfrak{I} \tilde{x}_{k}\right)^2  
\end{equation}
which, by definition, follows a chi-squared distribution with \emph{two} degrees of
freedom $\tilde\rho \sim \chi^2_{2}$.
This same quantity is referred to as $2\F_{2}$ in \cite{Dreissigacker:2018afk}.

The case of a chi-squared distribution with two degrees of freedom is degenerate 
with an exponential distribution. For the sake of clarity, we re-express it as 
a gamma distribution with shape parameter $k=1$ and scale parameter $\theta = 2$, i.e. 
$\tilde\rho \sim \Gamma(1, 2)$. We note that chi-squared distributions correspond to the locus 
\mbox{$\theta=2$} in the parameter space of Gamma distributions, with $k$ equal to half the degrees
of freedom; exponential distributions correspond to the locus \mbox{$k = 1$}, with $\theta$ equal
to the inverse of the rate parameter.

Let us now define $x_{n}^{p}$ as the zero-padded time series containing $N p$ elements, the 
last $N(p - 1)$ of which are purposely zero. This padding re-scales the variance of the original 
distribution by a factor $1/p$ and the resulting power can be expressed as 
\begin{equation}
    \tilde\rho_{p}
    = \left(\sqrt{\frac{2p}{N}}\:\mathfrak{R} \tilde{x}^{p}_{k}\right)^2  
    + \left(\sqrt{\frac{2p}{N}}\:\mathfrak{I} \tilde{x}^{p}_{k}\right)^2  
    = p \tilde\rho\;.
    \label{eq:rho_p}
\end{equation}
Then, by the properties of the Gamma function, \mbox{$\tilde\rho_{p} \sim \Gamma(1, 2p)$}, which is 
\emph{not} a chi-squared distribution for $p > 1$, but an exponential distribution with rate parameter 
$\lambda=(2p)^{-1}$.

Finally, we discuss the asymptotics of the distribution followed by the maximum of a 
\mbox{$\Gamma$-distributed} random variable.
As explained in Sec.~\ref{subsec:Noise}, such \mbox{light-tailed} distributions fall under the 
domain of attraction of the Gumbel distribution, meaning
\begin{equation}
    \max_{\N}{\Gamma(k, \theta}) \xrightarrow{\N \to \infty} \textrm{Gumbel}(\mu, \sigma)\;,
    \label{eq:mu_gamma}
\end{equation}
where the location and scale parameters $(\mu, \sigma)$ are given by  \cite{embrechts2013modelling}
\begin{equation}
    \mu = \theta \left[\ln{\N} + (k-1) \ln{\ln{\N}} - \ln{\Gamma(k)}\right]\;,
\end{equation}
\begin{equation}
    \sigma = \theta\;.
    \label{eq:sigma_gamma}
\end{equation}
In particular, the case of \mbox{$\max_{\N}{\tilde\rho_{p}}$} results in
\begin{equation}
    \mu_{p}(\N) = \updated{2}p \ln{\N}\;,\; \sigma_{p} = 2p \;.
    \label{eq:padding_gumbel}
\end{equation}

It is clear from Eq.~\eqref{eq:sigma_gamma} that the asymptotic distribution described by
Eq.~\eqref{eq:prob_max} is a Gumbel distribution with a scale parameter $\sigma = 2$. On the 
other hand, the asymptotic distribution followed by \mbox{zero-padded} Gaussian noise ($p > 1$)
follows a Gumbel distribution with a scale factor \mbox{$\sigma_{p} = 2p > 2$}. Since the scale
parameter is \emph{independent} of $\N$, Eq.~\eqref{eq:prob_max} \emph{fails} to describe the
asymptotic distribution stemming from correlated templates. 
In other words, parameter-space correlations shift the distribution followed by the power
statistic away from the locus of chi-squared distributions; since these correlations are
generally contained in a certain characteristic length, the resulting light tails
are still, however, within the Gumbel distribution's domain of attraction 
\cite{leadbetter1983extremes}.

\begin{figure*}
    \includegraphics[width=\columnwidth]{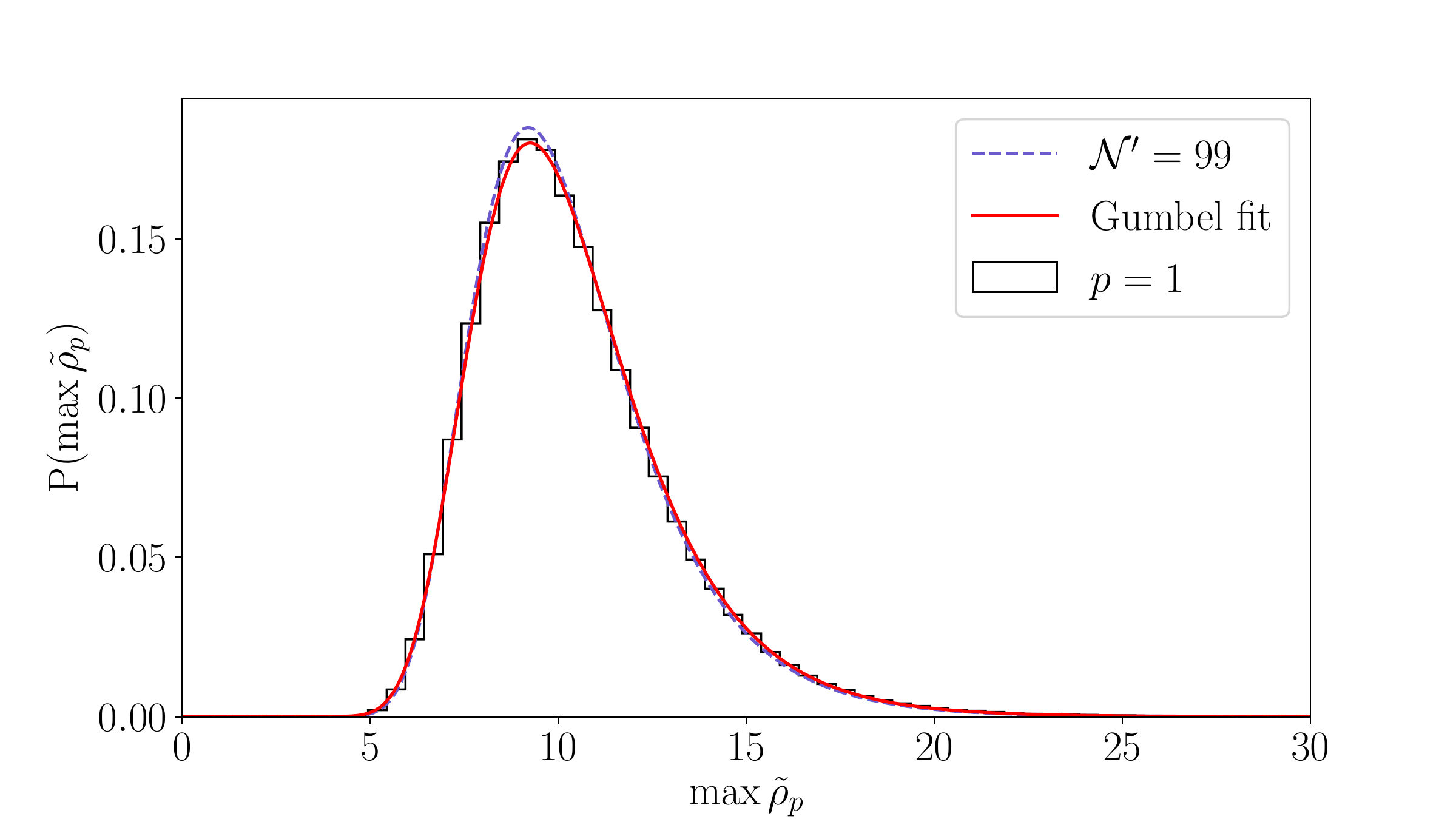}
    \includegraphics[width=\columnwidth]{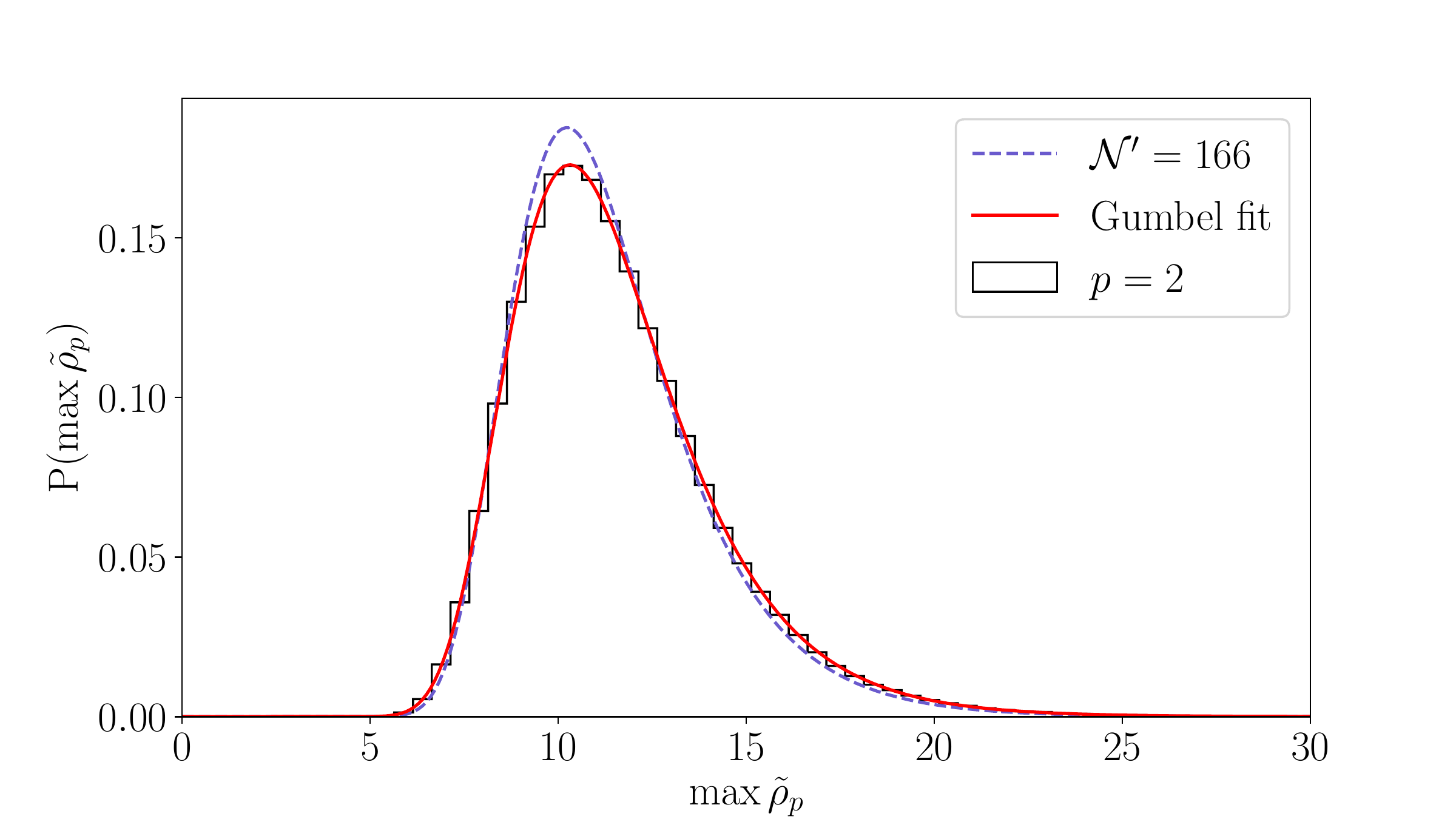}
    \includegraphics[width=\columnwidth]{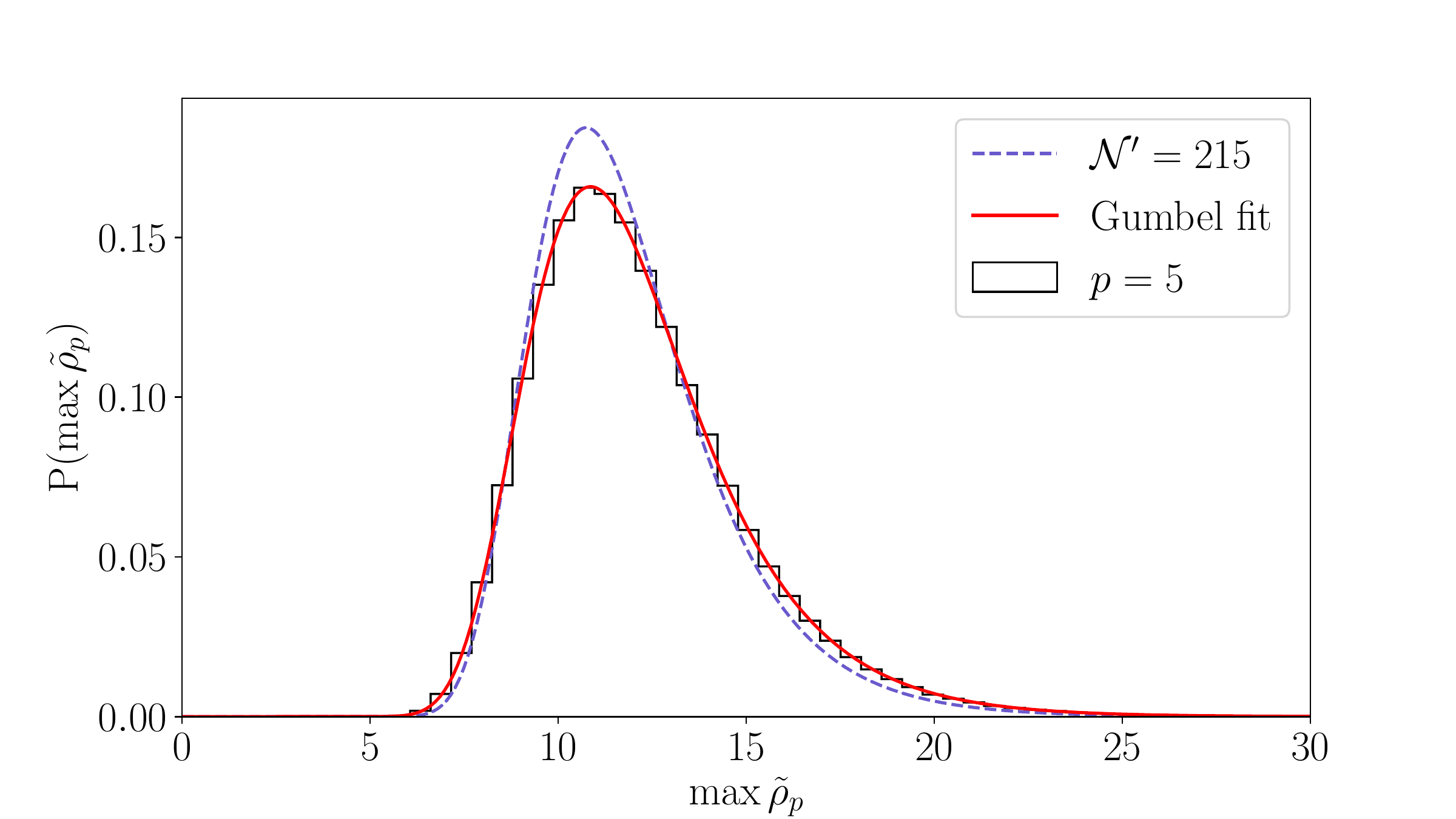}
    \includegraphics[width=\columnwidth]{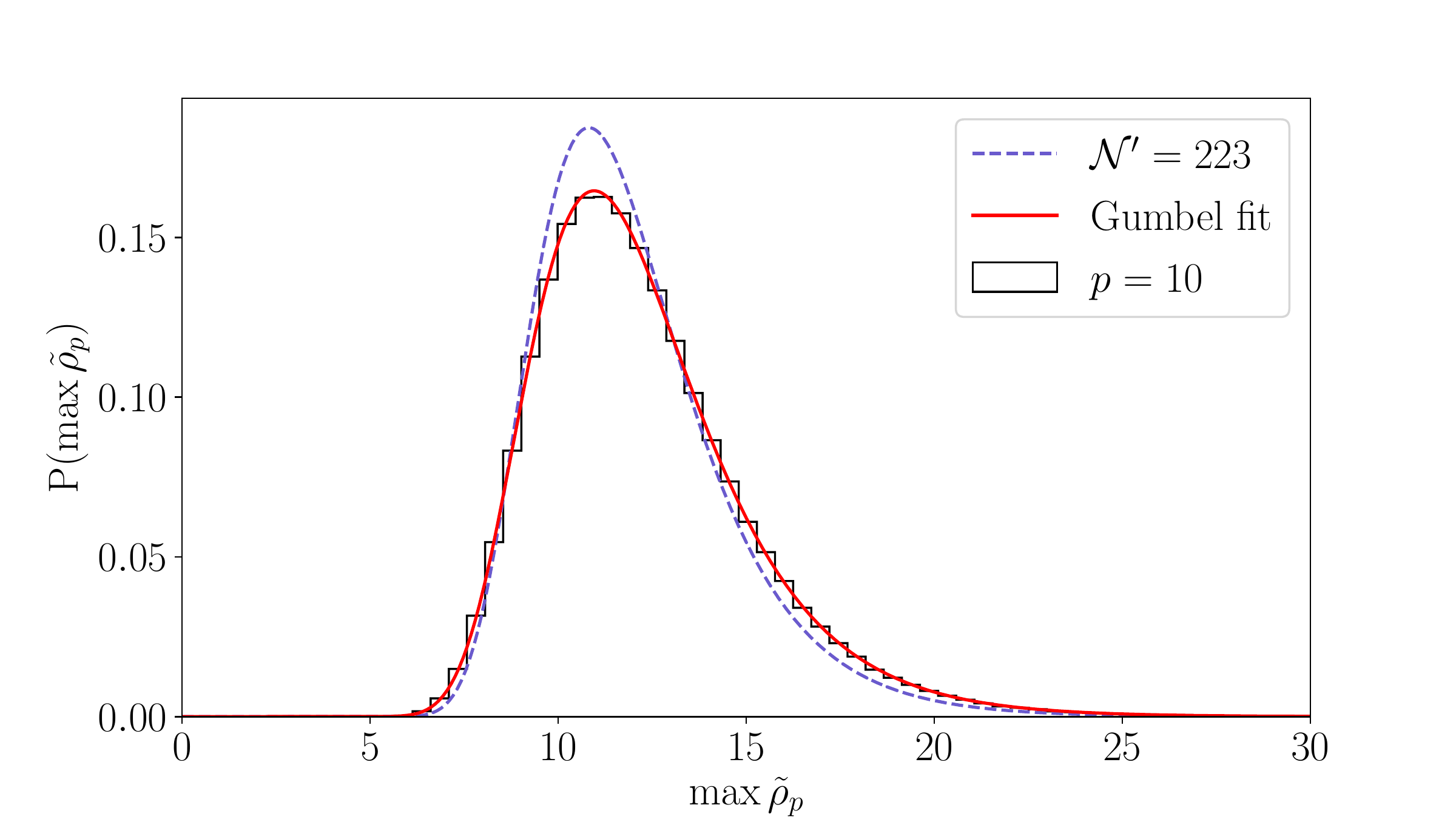}
    \caption{Maximum Fourier power over $N=200$ samples of zero-mean unit-variance zero-padded Gaussian noise.
    In each panel, the stair-case line represents a histogram over $10^6$ repeated trials of 
    $\max_{\N}{\tilde\rho_{p}}$. The dashed line is the best fit of Eq.~\eqref{eq:prob_max} on the effective 
    number of templates $\N'$, and the solid line is the best fit of a Gumbel distribution on the location and 
    scale parameters. Zero-padding is indicated by $p$, where $p=1$ represents no zero-padding, as explained 
    in the text.}
    \label{fig:oversampling}
\end{figure*}

This result is consistent with the findings reported in \mbox{Fig. 11} of \cite{Dreissigacker:2018afk},
which we reproduce in Fig.~\ref{fig:oversampling}. As the zero-padding increases, $\sigma_p$ increases 
and the resulting distribution, which is well described by a Gumbel distribution, spreads beyond the 
fit provided by Eq.~\eqref{eq:prob_max}.

\begin{figure*}
    \includegraphics[width=\columnwidth]{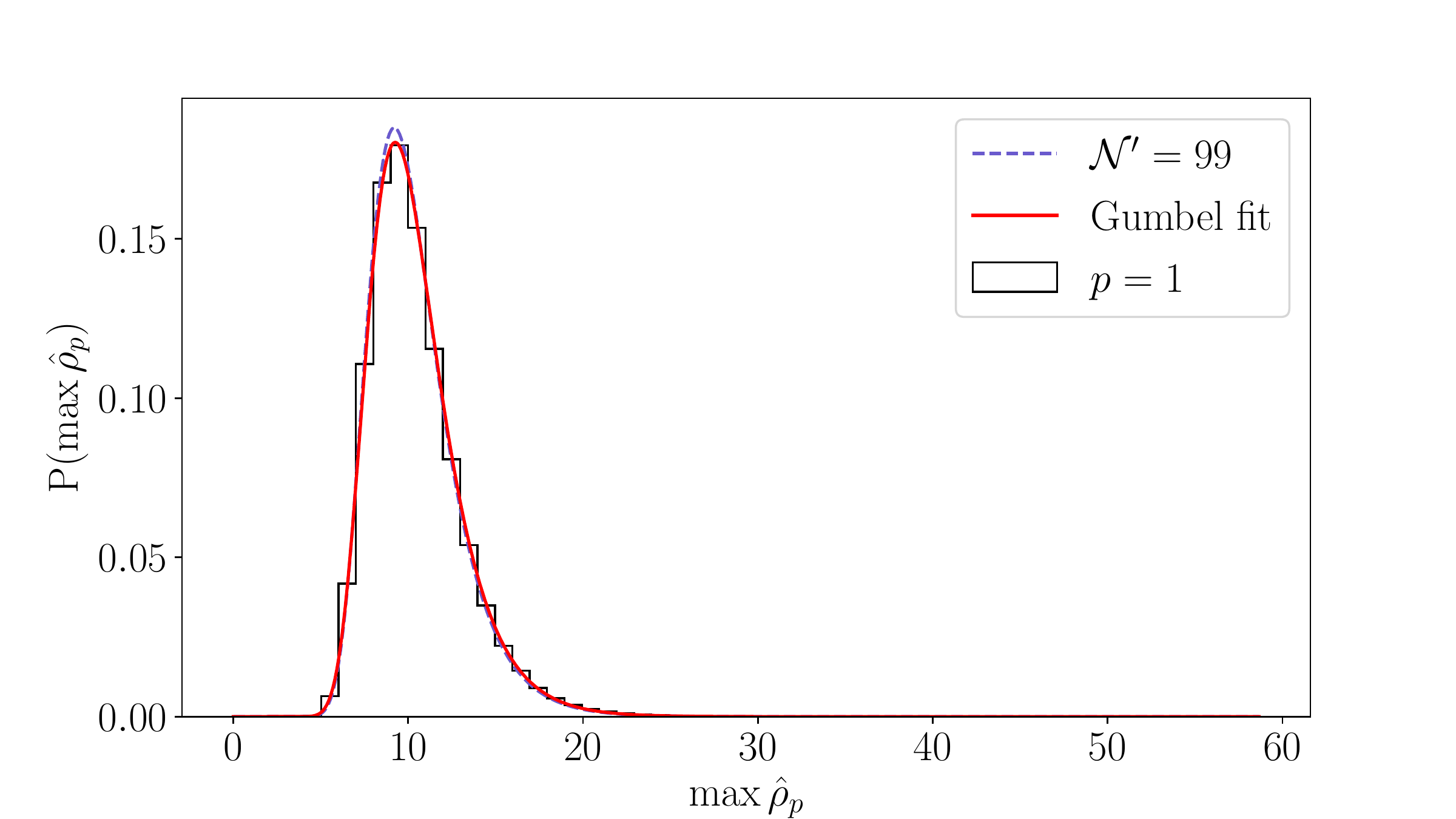}
    \includegraphics[width=\columnwidth]{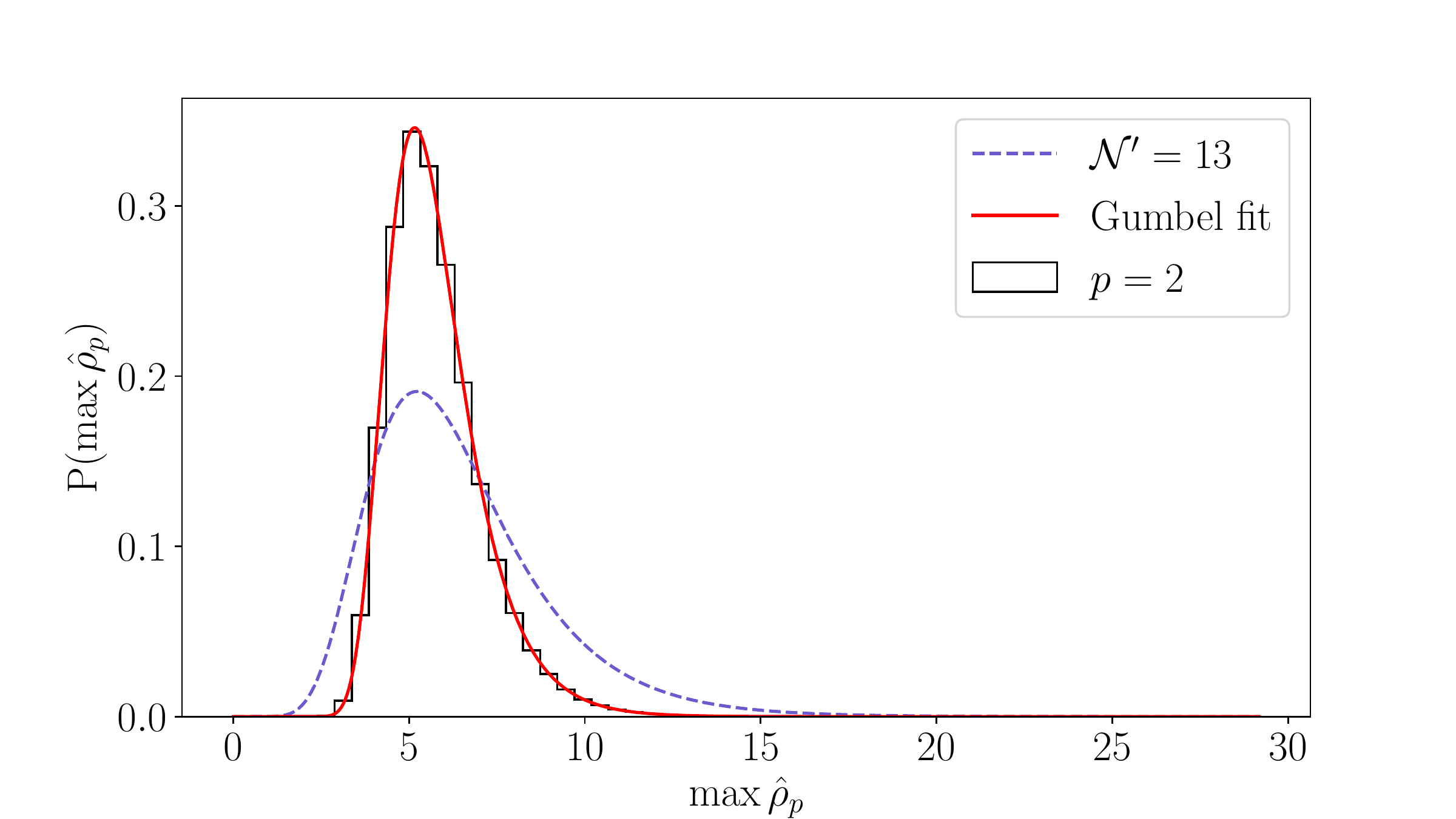}
    \caption{Equivalent figure to Fig. \ref{fig:oversampling} using the alternative normalization of Fourier power
    $\hat\rho_{p}$. In this case, increasing the number of correlated templates \emph{narrows} the 
    resulting distribution with respect to Eq.~\eqref{eq:prob_max}.}
    \label{fig:oversampling_obvious}
\end{figure*}

The location parameter $\mu$, on the other hand, does depend on the number of templates.
Indeed, if one tries to compute the required effective number of templates $\N'$ so that $\mu_{p=1}(\N')$ 
coincides with $\mu_{p}(\N)$,
\updated{
\begin{equation}
    \mu_{p=1}(\N') = 2 \ln{\N'} = 2 p \ln{\N} = \mu_{p}(\N)\;,
\end{equation}
}
the result is
\begin{equation}
    \N' = \N^{p} \;,
\end{equation}
which is a monotonic function of $p$. As a result, the effective number of templates
\emph{increases} with the zero-padding factor, again in agreement with \cite{Dreissigacker:2018afk}.
We note, however, that this is just a consequence of the chosen Fourier normalization.
If the normalized power was constructed using $N p$ as a normalization (the actual number of samples)
rather than $N$ (the number of non-zero-padded samples), then Eq.~\eqref{eq:rho_p} would be re-written as
\begin{equation}
    \hat\rho_{p} = \frac{1}{p} \tilde \rho\;.
\end{equation}
Consequently \mbox{$\hat\rho_{p} \sim \Gamma(1, 2/p)$} and the effects on the standard deviation would be 
exactly the opposite, as shown in Fig.~\ref{fig:oversampling_obvious}. Indeed, in such a case the effective 
number of templates would be $\N' = \N^{1/p}$, which decreases as the zero-padding increases.

Our proposed solution to the problem of estimating the effective number of templates is then 
\emph{not to do so}, as it depends strongly on the specific distribution followed by the noise, which is
generally unknown in a real case. Instead, we propose to describe the background noise distribution by fitting
an extreme value distribution to a set of samples (see e.g. Section \ref{subsec:Noise}). For \mbox{light-tailed}
noise, the proper distribution is Gumbel; other distributions are available for noise falling off as a power law
or presenting an upper cut-off.

\bibliography{references}

\end{document}